\documentclass[aps,twocolumn,prx,amsmath,amssymb,floatfix,superscriptaddress,footnoteinbib]{revtex4-1}
\usepackage[english]{babel}
\usepackage{multirow}
\usepackage{graphicx}
\usepackage{bm}
\usepackage{slashed}
\usepackage{amsmath}
\usepackage{epstopdf}
\usepackage{amsfonts}
\usepackage{amssymb}%
\usepackage{xcolor}
\usepackage{ dsfont }

\renewcommand{\Re}{{\rm \, Re\,}}
\renewcommand{\Im}{{\rm \, Im\,}}

\renewcommand{\hat}[1]{{\widehat #1}}
\newcommand{\diag}{{\rm \, diag\,}}

\newcommand{\aleq}[1]{
\begin{equation}
    \begin{aligned}
    #1
    \end{aligned}
\end{equation}
}

\providecommand{\U}[1]{\protect\rule{.1in}{.1in}}

\providecommand{\U}[1]{\protect\rule{.1in}{.1in}}
\begin{document}
\title{Tunneling into a Luttinger liquid coupled to acoustic phonons out of equilibrium}
\author{P.A. Nosov}
\address{Stanford Institute for Theoretical Physics, Stanford 
University, Stanford, CA 94305, USA}

\author{R.A. Niyazov}
\address{Department of Physics, St. Petersburg State University, St. Petersburg 199034, Russia}
\address{NRC ``Kurchatov Institute'', Petersburg Nuclear Physics Institute, Gatchina 188300, Russia}

\author{D.N. Aristov}
\address{NRC ``Kurchatov Institute'', Petersburg Nuclear Physics Institute, Gatchina 188300, Russia}
\address{Department of Physics, St. Petersburg State University, St. Petersburg 199034, Russia}
\address{Institut fur Nanotechnologie, Karlsruhe Institute of Technology, 76021 Karlsruhe, Germany}


\begin{abstract}
The renormalization of conductances in a Y junction of spinless Luttinger-liquid wires additionally coupled to acoustic longitudinal phonons is investigated in fermionic representation. This system corresponds to geometry of a tunneling experiment and exhibits the interplay between the Coulomb repulsion and the attractive retarded interaction mediated by phonons. The retardation effects related to the propagation of phonons through the junction with arbitrary transmission and reflection amplitudes are taken into account. The appearing logarithmic corrections to conductances of the junction are treated in a renormalization group approach, and scaling exponents are calculated up to infinite order in the interaction after RPA-type summation. 
The fixed points and corresponding scaling exponents are considered in various non-equilibrium regimes. 
We show that the boundary exponent and the bulk anomalous dimension of fermion operator are characterized by two different Luttinger parameters, referring to the main wire, thanks to non-local character of phonon-mediated interaction.
In the limiting case of the junction of only two wires, the scaling exponents found by our method are in exact correspondence with previous bosonization analysis.
\end{abstract}

\pacs{Valid PACS appear here}
\maketitle

\section{Introduction}
\label{sec:intro}
One-dimensional quantum systems with electron-phonon interactions have been extensively studied in the literature \cite{Loss1994,Apostol1982} for their remarkable transport properties and practical implementations in carbon nanotube devices. It is also well known that embedding a potential impurity into the quantum wire with the repulsive electronic interactions leads to the power-law renormalization of its scattering amplitude and suppression of transparency, as was initially discussed within two complementary theoretical approaches, bosonization \cite{Kane1992} and conventional fermionic one \cite{Yue1994}.

Recently, the effect of electron-phonon interactions on the electrical conductance and transport properties of one-dimensional strongly correlated electronic systems was discussed in the context of helical edge states of two-dimensional topological insulators \cite{Budich2012,Groenendijk2018}, quantum Hall edge states at filling factor $\nu=1$ \cite{Idrisov2019} and topological insulator nanowires \cite{Dorn2020}. It was emphasised that preserving time-reversal symmetry inelastic scattering processes due to phonons can drastically influence the topologically protected transport properties. Such dissipative mechanisms induce backscattering in the presence of the Rashba spin orbit coupling \cite{Budich2012} or the spin-polarized tunneling tip \cite{Aristov_2017} which, in principle, might lead to significant corrections to measurable conductances or even to the existence of new fixed points in renormalization group sense.

The later possibility was considered in \cite{Galda2011} by means of the functional bosonization formalism. It was argued that a Luttinger Liquid (LL) with the electron-phonon interaction and a single impurity exhibits an intermediate state related to the new unstable fixed point in which the system can flow either to the metallic or insulating limit, depending on the impurity strength.

In the present paper, we further investigate the renormalization group structure of one-dimensional electron-phonon liquids and extend it to the non-equilibrium case with more complicated geometry.
Specifically, we estimate the conductance scaling of a Y junction of spinless Luttinger-liquid wires additionally coupled to acoustic longitudinal phonons as functions of bias voltages applied to three independent Fermi-liquid reservoirs. 

We adopt the fully-fermionic approach \cite{Yue1994}
with naturally incorporated thermal Fermi
reservoirs which allows us to avoid, by construction, difficulties 
arising in the bosonization technique
when interactions are considered only within a finite segment of
wires \cite{Maslov1995,Oshikawa2006,Wang2011}. Perturbative fermionic theory of RG, formulated in the paper \cite{Yue1994}, has been effectively used in a variety of problems, for instance for a double barrier \cite{Polyakov2003, Nazarov2003} and Y-junctions in LL
\cite{Siddhartha2002,Aristov2010, Affleck2016,Shi_2016}.

The system exhibits the interplay between the Coulomb repulsion and the attractive, retarded interaction mediated by the phonons. We take into account non-local effects that correspond to the propagation of phonons through the impurity with arbitrary transmission and reflection amplitudes. In these conditions, we at first analyze corrections to conductances due to the electron-phonon interaction at the one-loop level in the Keldysh formalism and treat the appearing logarithmic singularities by means of renormalization group scheme developed in \cite{Aristov2017}. Furthermore, we extend our results to
an arbitrary interaction strength by summing an infinite series in perturbation theory (RPA-type summation).

In particular, we obtain that the electron-phonon coupling drastically changes the phase diagram of the system. If the coupling parameters are sufficiently large then the relevant and irrelevant parts of the flow diagram interchange and the different ``metallic'' scaling behavior appears. Various non-equilibrium regimes are considered. Scaling exponents for conductances are calculated in all orders in the perturbation theory both in the electron-phonon and electron-electron interactions. In the limiting case of a 2-wire junction the scaling exponents found by our method are in exact correspondence with previous bosonization investigations \cite{Yurkevich2013}.

We also discuss how the boundary exponent and bulk anomalous dimension of the fermion operator are modified in the presence of non-local processes associated with phonons in a Y-junction geometry. Previous studies show \cite{Aristov2011} that in the presence of only local interactions it is possible to independently extract from both of these quantities a {\it single} Luttinger parameter which solely governs renormalization of conductances. In contrast to that, we demonstrate that in the tunneling experiments, where the phonons pass unhindered the vicinity of Y junction, this commonly accepted procedure is ill-defined and rather requires two effective Luttinger parameters.

The paper is organized as follows. In Sec.~\ref{sec:model} we formulate our model for a Y junction in the presence of fermion-phonon interactions. Non-equilibrium RG equations for conductances up to the first order of perturbation theory are discussed in Sec.~\ref{sec:III_first_order} along with our general RG formalism.
The RPA-type summation to infinite order in the interaction is described in Sec.~\ref{ladder_section}.
Section~\ref{sec:RG_eq_strong} is devoted to the derivation of RG equations at strong coupling out of equilibrium. The solution of RG equations at strong coupling is presented in
Sec.~\ref{sec:scaling_exp_strong}. Sec.~\ref{sec:diss} is reserved for conclusion.

\section{The model}
\label{sec:model}
\subsection{Scattering state description of one-dimensional fermions}
\begin{figure}[h]
\includegraphics[width=0.8\columnwidth]{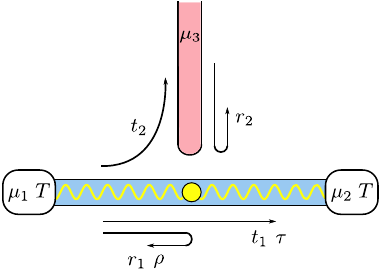}
\caption{{Setup of the Y-junction between a main wire (horizontal, blue) and a tunneling tip (vertical, red). The yellow wave represents the elastic degrees of freedom (acoustic phonons). The arrows correspond to the matrix elements of electrons and phonons scattering processes.}}
\label{Fig:Setup}
\end{figure}

We consider the following setup (see Fig.~\ref{Fig:Setup}): a system of spinless fermions in one dimension, interacting via a short-ranged screened Coulomb interaction in each of 3 quantum wires in the
regions $\ell < x_j < L$, $j=1,2,3$. These regions are assumed to be adiabatically connected to the Fermi-liquid leads at $x_j>L$. 

There is a junction in the narrow region $|x| < \ell$ , which scatters the fermions as described by the unitary $S$-matrix (up to overall phase factors in the individual wires).
 \begin{equation}\label{smatrix}
    S=
        \begin{pmatrix}
        r_1 & t_1 & t_2\\
        t_1 & r_1 & t_2\\
        t_2 & t_2 & r_2
        \end{pmatrix} \,.
    \end{equation}
Currents flowing towards a junction, $I_i$, and the chemical potentials, $\mu_j$, are related by $I_i=G_{ij} \mu_j$, with  the conductances matrix, $G_{ij}$.

In the absence of interactions this quantity is connected with the $S$ matrix by the Landauer formula $G_{ij}= \delta_{ij}-|S_{ij}|^2$. It follows from this formula that Kirchhoff's rules are obeyed due to unitarity of $S$ matrix. The interactions lead to renormalization of the conductances which is of the main interest in this paper.
    
We study interacting fermions in Tomonaga Luttinger model, described by the  Hamiltonian 
 \begin{equation}
\begin{aligned} 
&\mathcal{H}_{\text{LL}} =\int\limits_{0}^{\infty}dx \sum\limits_{j=1}^{3}\left[H_j^{0}(x)+H_j^{int}(x)\Theta(x;\ell,L) \right] \,,  \\ 
& H_j^0(x) =v\left[\psi^\dagger_{j,in}(x) i\nabla \psi_{j,in}(x)-\psi^\dagger_{j,out}(x) i\nabla \psi_{j,out}(x) \right] 
 \,, \\
& H_j^{int}(x)   =2\pi v g_{j} \psi^\dagger_{j,in}(x) \psi_{j,in}(x)\psi^\dagger_{j,out}(x) \psi_{j,out}(x)  \,.  \\
\end{aligned}
\end{equation}
Here $v$ is the Fermi velocity, $g_j$ is the interaction constant in the lead $j$ and $\Theta(x;\ell,L)= 1$ in the interval $\ell <|x| < L$ and zero elsewhere. The fermionic field operators $\psi^\dagger_{j,\eta_j}(x)$ create particles at position $x$ in scattering states
$|j, \eta_j ; \omega \rangle$ of energy $\omega$, in wire $j$ and with chirality $\eta_j = \pm 1$,
labeling incoming ($\eta_j = - 1$) and outgoing ($\eta_j = + 1$) states. For simplicity we use the compact definition $j_\eta=(j,\eta_j)$.
The outgoing fermion operators are connected with the
incoming ones by the $S$ matrix, $\psi_{j,out}(0) = S_{jk}\psi_{k,in}(0)$.

\subsection{Coupling to the acoustic phonons}
In this paper we consider fermions   coupled to one-dimensional acoustic phonons. We assume that the phonon spectrum is linear up to a cutoff at the Debye energy $\omega_D$. Phonons are linearly coupled to the electron density and the electron-phonon interaction takes place within the same region $\ell < x_j < L$ in each wire  and are described by the following Hamiltonian 
 \begin{equation}
\begin{aligned}
&\mathcal{H}_{\text{ph}}  =\int\limits_{0}^{\infty}dx \sum\limits_{j=1}^{3}\left[H^{0,\text{ph}}_{j}(x)+H^{\text{el-ph}}_{j}(x)\Theta(x;\ell,L) \right] \,, \\ 
&H^{0,\text{ph}}_{j}(x) =\left[\frac{1}{2}\dot{u}_j^2(x)+\frac{1}{2}(c\nabla u_j(x))^2\right]  \, ,\\
&H^{\text{el-ph}}_{j}(x) =\sqrt{\pi v \alpha_j} c\nabla u_j(x)\sum\limits_{\eta=\pm 1} \psi^\dagger_{j,\eta}(x) \psi_{j,\eta}(x) \,.    \\
\end{aligned}
\end{equation}

 Here $c$ is the  speed of sound, $\alpha_j$ is a dimensionless electron-phonon coupling constant. The phonons are also scattered by the vicinity of the junction, which is encoded below in the Green's function for the displacement operator $u_j(x)$ in the $j$-wire,  see Appendix~\ref{AppendixA} for more details. 
The effective electron-electron interaction via phonons, with the energy transfer below the Debye frequency $|\omega| \alt\omega_D$, takes the form
\begin{multline} \label{eq:greenphon}
   L^{R,(0)}_{\alpha,\omega} (l_\eta,x|m_\eta,y)=  \pi v \sqrt{\alpha_l \alpha_m}
\langle  c\nabla u_l(x)  c\nabla u_m(y)\rangle^{ret}_\omega   \,, \\
= - \pi v \alpha_l\delta_{lm}\delta(x-y) 
     \phantom{-}\\
   - \frac{i\omega \pi v }{2c}\sqrt{\alpha_l \alpha_m}\left( e^{i\frac{\omega}{c} |x-y|} \delta_{lm} + e^{i\frac{\omega}{c} (x+y)} B_{lm}\right) \,,
\end{multline}
where the retarded Green's function for the gradients of deformations was obtained from \eqref{GF_deformations}. The matrix $\mathbf{B}$  strongly influences the results for the electrical conductance and is some analog of the squared scattering matrix $|S_{ij}|^2$.

We focus on the symmetric Y-junction geometry which corresponds to a tunneling experiment: the electron-phonon interaction is present only in two wires (which together form a ``main wire''). Another wire (``tunneling tip'') contains only the electron-electron interaction and the role of phonons is negligible there, it implies the absence of   phonon transport between the main wire and the tunneling tip. All these conventions can be encoded by writing $\alpha_j=\alpha(1-\delta_{j3})$ and by the following form of the $\mathbf{B}$ matrix 
\begin{equation} \label{eq:Bmat}
\mathbf{B}=
        \begin{pmatrix}
        \rho & \tau & 0\\
         \tau & \rho & 0\\
        0&0&1
        \end{pmatrix}  \,.
\end{equation}
We assume below that the phonon transmission, $\tau$, and reflection, $\rho$, coefficients to be constrained by the ``unitarity'' condition $\tau+\rho=1$ (see Appendix~\ref{AppendixA}), which will significantly simplify our calculations. However, our formalism is not restricted to this case and more general form of \eqref{eq:Bmat} can be used as well.

\begin{figure}
\includegraphics[width=0.6\columnwidth]{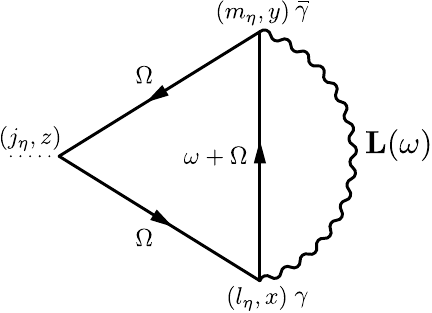}
\caption{The diagram leading to the current correction due to interaction. Wavy line is the sum of bare Coulomb  and electron-phonon interactions in case of first-order correction computation. 
}
\label{Fig:Cond}
\end{figure}

\section{First order corrections to conductances}
\label{sec:III_first_order}

\subsection{General formalism}
We employ the fermionic approach for the calculation of interaction-induced corrections to the currents. It was developed in the paper~\cite{Aristov2014} by using the Keldysh technique for steady-current, off-equilibrium case of the leads maintained at non-equal chemical potentials . 
The first-order corrections to the currents  are  described  by  the  diagram  in  Fig.~\ref{Fig:Cond}. Solid lines correspond to the fermions Green functions which in Keldysh space (denoted by an underbar) have a structure 
 \begin{equation}
\underline{G} = 
    \begin{pmatrix}
        G^R & G^K \\
        0 & G^A\\
    \end{pmatrix}.
 \end{equation}
The Green's function in chirality index space have a structure of $2\times2$ matrix denoted by  square brackets and  a hat $\hat G_{\eta_l \eta_j} (l,y|j,x)= G(l,\eta_l,y|j,\eta_j, x)$
\begin{equation}\label{GF_def}
    \begin{aligned}{}
    \hat G^R_\omega(l,y|j,x)&=-\frac{i}{v} \theta(\tau) e^{i \omega \tau} 
    \begin{bmatrix}
    \delta_{lj} & 0\\
    S_{lj} & \delta_{lj}\\
    \end{bmatrix}, \\
    \hat G^A_\omega(l,y|j,x)&=\frac{i}{v} \theta(-\tau) e^{i \omega \tau} 
    \begin{bmatrix}
    \delta_{lj} & S_{jl}^*\\
    0 & \delta_{lj}\\
    \end{bmatrix}, \\
    \hat G^K_\omega(l,y|j,x)&=-\frac{i}{v}  e^{i \omega \tau} 
    \begin{bmatrix}
    \delta_{lj} h_l &  S_{jl}^* h_l\\
    S_{lj} h_j & S_{jm}^* S_{lm} h_m\\
    \end{bmatrix}, \\
    \tau&=(\eta_l y-\eta x_j)/v.
    \end{aligned}
\end{equation}
Here, $h_j(\omega) = \text{tanh}[(\omega - \mu_j)/2T]$ is the equilibrium distribution function in the  lead $j$ with the chemical potential $\mu_j$. 

The wavy line in Fig.\ \ref{Fig:Cond} denotes the full interaction between fermions in the system. It is diagonal in  Keldysh space and originates from two contributions. The first one is the electron-electron part of interaction 
\begin{equation}\label{eq:elel0}
    \hat L^{(0)}_{g,\omega} (l,x|m,y)= 2 \pi v g_l \delta_{lm} \delta(x-y) 
    \begin{bmatrix}
    1 & 1\\
    1 & 1\\
    \end{bmatrix}.
\end{equation}
We consider the junction of a main wire and a tunneling tip so that we put $g_1=g_2=g$ with  $g_3 \neq g$. The second contribution is the electron-phonon interaction with the retarded component~\eqref{eq:greenphon}. Combining \eqref{eq:elel0} and \eqref{eq:greenphon} together, we obtain the full interaction propagator for symmetric Y-junction geometry in the following form
\aleq{\label{L0_bare}
   L^{R,(0)}_{\omega} &(l_\eta,x|m_\eta,y) =L^{R,(0)}_{g,\omega}(l_\eta,x|m_\eta,y)\\ +&L^{R,(0)}_{\alpha,\omega}(l_\eta,x|m_\eta,y)
 = \lambda_l\delta\left(x-y\right)\delta_{lm}\\
    +&  i\omega \zeta \left( e^{i\frac{\omega}{c} |x-y|} \delta_{lm}(1-\delta_{l3})+e^{i\frac{\omega}{c} (x+y)} B_{lm} \right), 
}
where $\lambda_{1,2}=2\pi v (g-\alpha/2),$  $\lambda_3=2\pi v g_3,$  and $\zeta=-\pi\alpha v/2c$. 
The parameter  $\xi =v/c$ is the ratio of plasmon and phonon velocities.
The advanced component of interaction propagator is given by 
\begin{equation}
    \hat L^A=(\hat L^R)^\dagger|_{x\leftrightarrow y}.
\end{equation}
We emphasize that  the Keldysh components of interactions,  in fact, do not produce logarithmic corrections to currents \cite{Aristov2014} due to the dominant role of virtual  processes in renormalization. Moreover, retarded and Keldysh components do not mix with each other upon the RPA-type summation of higher order diagrams allowing us to concentrate henceforth only on the retarded component of the interaction. 

Let us now briefly remind a way to calculate a one-loop correction to the currents. The corresponding diagram is depicted in Fig.~\ref{Fig:Cond} and can be expressed as
\begin{equation}\label{eq:curr1}
    \begin{aligned}
        J_{j_\eta}^{(1)}(z)=&i\int \frac{d\omega}{2 \pi}\int dx dy \sum_{l_\eta, m_\eta} \\
        &\times \text{Tr}_K[\underline{T}_{\omega} (m_\eta , y| l_\eta,x;j_\eta , z) \underline{L}^{(0)}_\omega(l_\eta,x|m_\eta,y)],
    \end{aligned}
\end{equation}
where we sum over all wire indices, chirality components, and $\text{Tr}_K$ is the trace over the Keldysh indices. In addition, we introduced the structural part $T$ of the diagram corresponding to the triangle of fermion Green's function
\begin{equation}\label{eq:T}
   \begin{aligned}
   T^{\nu \mu}_\omega &(m_\eta,y|l_\eta,x;j_\eta,z) \\
   =& v_j \int \frac{d\Omega}{2 \pi}  \text{Tr}_K[\underline{\gamma_{\text{ext}}} \underline{\hat{G}}_\Omega (j_\eta,z|m_\eta,y) \underline{\bar{\gamma}}^\nu \\
   &\times \underline{\hat{G}}_{\Omega+\omega}(m_\eta,y|l_\eta,x)  \underline{\gamma}^\mu\underline{\hat{G}}_\Omega(l_\eta,x|j_\eta,z)].
   \end{aligned}
\end{equation}
This fermionic triangle $T$ turns out to be a function of two external points with different wire indices and coordinates due to the presence of the retarded and matrix parts of the interaction \eqref{eq:greenphon} (in contrast to the case of the screened Coulomb interaction out of equilibrium previously discussed in the literature  \cite{Aristov2017}).

External and interaction vertices are defined as
\aleq{
\underline{\gamma}_\text{ext} =& \frac{i}{2}
    \begin{pmatrix}
        1 & 1 \\
        -1 & -1\\
    \end{pmatrix}, \\
\underline{\gamma}^1 =\underline{\bar{\gamma}}^2 =& \frac{1}{\sqrt{2}}
    \begin{pmatrix}
        1 & 0 \\
        0 & 1\\
    \end{pmatrix},  \\
\underline{\gamma}^2 = \underline{\bar{\gamma}}^1 =& \frac{1}{\sqrt{2}}
    \begin{pmatrix}
        0 & 1 \\
        1 & 0\\
    \end{pmatrix}, 
}

The diagram in Fig.~\ref{Fig:Cond} should be combined with the one where the arrows of the fermionic lines are reverted.  

Integration over $\Omega$ leads to two generic integrals:
\aleq{
\int d\Omega [h_j (\Omega+ \omega) - h_j (\Omega)]&=2 \omega, \\
\int d\Omega [1-h_j (\Omega+ \omega) h_p (\Omega)]&=2 f_2(\omega+\mu_p-\mu_j),
}
where $f_2(x)=x \coth{(x/2T)}$.
Calculation shows that $T_{21}=0$ so that the Keldysh component of   interaction does not contribute to the current~\eqref{eq:curr1} as is mentioned above.

We assume that the  point $z$ lies outside the interaction region. In this case the dependence on $z$  in outgoing current $J_{j_+}^{(1)}(z>L) \equiv J_{j}^{(1)}$ disappears, whereas correction to the incoming current is absent $J_{j_-}^{(1)}(z>L)=0$. One can verify the charge conservation law $\sum_{j} J_{j}^{(1)}=0$. 

 By explicitly evaluating all matrix elements of \eqref{eq:T} one can show that the infrared(IR) regularized correction to the current reads
\begin{equation}\label{J_full_equation}
     J_{j}^{(1)}=- \Im \sum_{mkp} M^j_{mkp}  \int\limits_{\epsilon}^{\omega_0} \frac{d\omega}{4\pi v^2} L^{(0)}_{\omega,\text{odd}}(m|k) F(\omega,V_{mp}) 
\end{equation}
with 
 \begin{equation}
\begin{aligned}
 F(\omega,V)&= (f_2(\omega+V)-f_2(\omega-V)) ,\\
 L^{R,(0)}_{\omega}(m|k)&=\int dxdy L^{R,(0)}_\omega(m_-,x|k_+,y) e^{i\frac{\omega}{v}(x+y)} ,\\
 L^{(0)}_{\omega,\text{odd}}(m|k)&= (L^{R,(0)}_{\omega}(m|k)-L^{R,(0)}_{-\omega}(m|k)), \\
   M^j_{mkp}& =S_{km}S^*_{j m} S^*_{kp} S_{j p}  \,.
\end{aligned}
\label{eq:L_simple_1}
\end{equation}
Here $\omega_0=v/\ell$ is the ultraviolet (UV) cut-off,  $\epsilon$ is the running IR energy scale, and  $V_{mp}=\mu_m-\mu_p$ is voltage between the wires $m$ and $p$.  

We stress that there is no contribution in \eqref{J_full_equation} associated with identical chiralities at both vertices (compare with \eqref{eq:L_simple_1}), despite the fact that the electron-phonon interaction is independent of chirality and formally contains corresponding nonzero matrix elements. Similarly to standard ``g-ology'',  it just means that the forward scattering small-momentum transfer amplitudes of the phonon-mediated and screened Coulomb interactions
do not participate in the renormalization of conductances in the dc limit. 

\subsection{First order of the interaction}  \label{sec:first_order}

The quantity $ L^{(0)}_{\omega,\text{odd}}$, defined in \eqref{eq:L_simple_1} and calculated for the interaction
\eqref{L0_bare}, contains several terms of the form   $\omega^{-1} \sin^{2} \omega t_{0}$ with  $t_{0}=  L/v,  L(1+\xi)/v, 2 L(1+\xi)/v$. The existence of these rapid oscillations ensures convergence of \eqref{J_full_equation}
at small $\omega$ and at the same time all such  terms can   be approximated by $(2\omega)^{-1}$ at larger energies,  $\omega\agt t_{0}^{-1} \sim L^{-1} $. Hence, in the limit $L \to \infty$ we can write
\aleq{ 
L^{(0)}_{\omega,\text{odd}}(m|k)=\frac{ 2\pi i v^2}{\omega}\left( (g-\bar\alpha(1+\xi))\delta_{mk}+\bar\alpha\xi B_{mk}\right),
}
where $\bar\alpha=\frac{\alpha}{2(1+\xi)^2}$.
 
The remaining integral over energy  is logarithmically divergent: 
\aleq{\label{log_integral}
\mathcal{I}&\left(\omega_{0}, \epsilon, V\right)=\int_{\epsilon}^{\omega_{0}} \frac{d \omega}{\omega} F(\omega, V)=2 V\theta(\epsilon-|V|)  \ln \frac{\omega_{0}}{\epsilon} \\ &+2 \theta(|V|-\epsilon)\left[V-\epsilon \operatorname{sgn}(V)+V \ln \frac{\omega_{0}}{V}\right] \\ &\simeq 2 V \;\ln \left(\frac{\omega_{0}}{\rm{max}\{|V|,\epsilon\}}\right) \,.
}
In the last line we neglected the contribution  from the region $ \epsilon <| V | $ because its dependence on the running scale $ \epsilon $ is only linear.

The upper limit of integration in \eqref{log_integral} should be used with caution: in processes involving phonons the UV cut-off $\omega_0$ should be replaced by the characteristic energy scale $\omega_D$. From the RG point of view, it just implies that all divergent logarithmic contributions proportional to the electron-phonon coupling constant $\alpha$ should be accompanied by the step function  $\theta_D(\epsilon)=\theta(\omega_D-\epsilon)$.

Summarizing, the first order correction to the currents reads as

\aleq{\label{Current_full_log_first_order}
       J_j^{(1)} =-\sum\limits_{mkp}\left( (g-\bar\alpha(1+\xi))\delta_{mk}+\bar\alpha\xi B_{mk}\right)\\\times  V_{mp}\operatorname{Re}[M_{mkp}^j]\; \ln \left(\frac{\omega_{0}}{\rm{max}\{|V_{mp}|,\epsilon\}} \right)\,.
}

It is convenient to introduce two independent currents
$J_{a,b}$ and two independent bias voltages $V_{a,b}$ as follows:
\begin{equation}
J_a=\frac{1}{2}\left(J_1-J_2 \right), \quad V_a=\mu_1-\mu_2
\end{equation}
for the main wire and
\aleq{
J_b=&\frac{1}{3}\left(J_1+J_2-2J_3 \right)=-J_3, \\
\quad V_b=&\frac{1}{2}\left(\mu_1+\mu_2 - 2 \mu_3 \right)
}
for the tunneling tip. The bare (differential) conductances are then defined as
\begin{equation}\label{currents_conductances}
   G_a= \partial J_a/\partial  V_a  ,\quad  G_b = \partial J_b/ \partial V_b \,.
\end{equation}
For  the symmetric $S$ matrix, Eq.\ \eqref{smatrix}, the bare conductances in non-interacting situation are
\begin{equation}\label{eq:barecond}
 G_a^{(0)}= \frac{1}{2}(1-|r_1|^2+|t_1|^2) ,\quad   G_b^{(0)}= 2 |t_2|^2\,.
\end{equation}
Without a loss of generality, we can assume $\mu_3=0$ and $V_{a,b}\geqslant 0$, so $|\mu_2|\leqslant \mu_1$. The unitarity of the $S$-matrix constrains the domain of allowed conductances in the $(G_a,G_b)$ plane by the straight line and the parabola $0\leq G_b \leq 4(G_a - G_a^2)$ \cite{Aristov2010}.

Having obtained the corrections to conductances from Eqs.\ \eqref{Current_full_log_first_order} and \eqref{currents_conductances}, and assuming the scaling behavior of $G_{a,b}$ (see Ref.~\cite{Aristov2017}) we can differentiate $G_{a,b}$ with respect to $\Lambda=\ln (\omega_0/\epsilon)$ in order to obtain a set of perturbative RG equations:
\aleq{\label{RGeq1g}
   \frac{d G_a}{d\Lambda} &= 2A_1 \theta_a(\epsilon)+A_2 \theta_+(\epsilon),\\
    \frac{d G_b}{d\Lambda} &= 2B_2 \theta_+(\epsilon),
}
with 
\begin{equation}
\begin{aligned}\label{A,B_weak}
A_1=&-\Big(g-\bar\alpha(1+2\xi \tau)\theta_D(\epsilon)\Big)\left(G_a(1-G_a)-\frac{G_b}{4}\right),\\
A_2=&-\frac{1}{8}\Big(g- \bar\alpha(1+\xi\tau)\theta_D(\epsilon)\Big) (1-G_a)G_b\\
&+\frac{\bar\alpha \xi\tau}{8}G_aG_b\theta_D(\epsilon) -\frac{g_3}{8}(1-2G_a)G_b,\\
B_2=&-\frac{1}{8}\Big(g-\bar\alpha (1+\xi\tau)\theta_D(\epsilon)\Big)\left[2-2G_a-G_b\right]G_b
\\&-\frac{g_3}{4}(1-G_b)G_b+\frac{\bar\alpha \xi\tau}{8 }\theta_D(\epsilon)(G_b-2G_a)G_b\,.
    \end{aligned}
\end{equation}
Here,  energy scales related to voltages are defined by Heaviside $\theta$-functions,  $\theta_a(\epsilon)=\theta(\epsilon-V_a)$ and $\theta_+(\epsilon)=\theta(\epsilon-\mu_2)+\theta(\epsilon-\mu_1)$. It means that the renormalization occurs in several steps with different right hand side of \eqref{RGeq1g} at each step. 
The phonon energy scale appears in  $\theta_D(\varepsilon)=\theta(\omega_D - \varepsilon)$.

\begin{figure*}
{\includegraphics[width=0.32\linewidth]{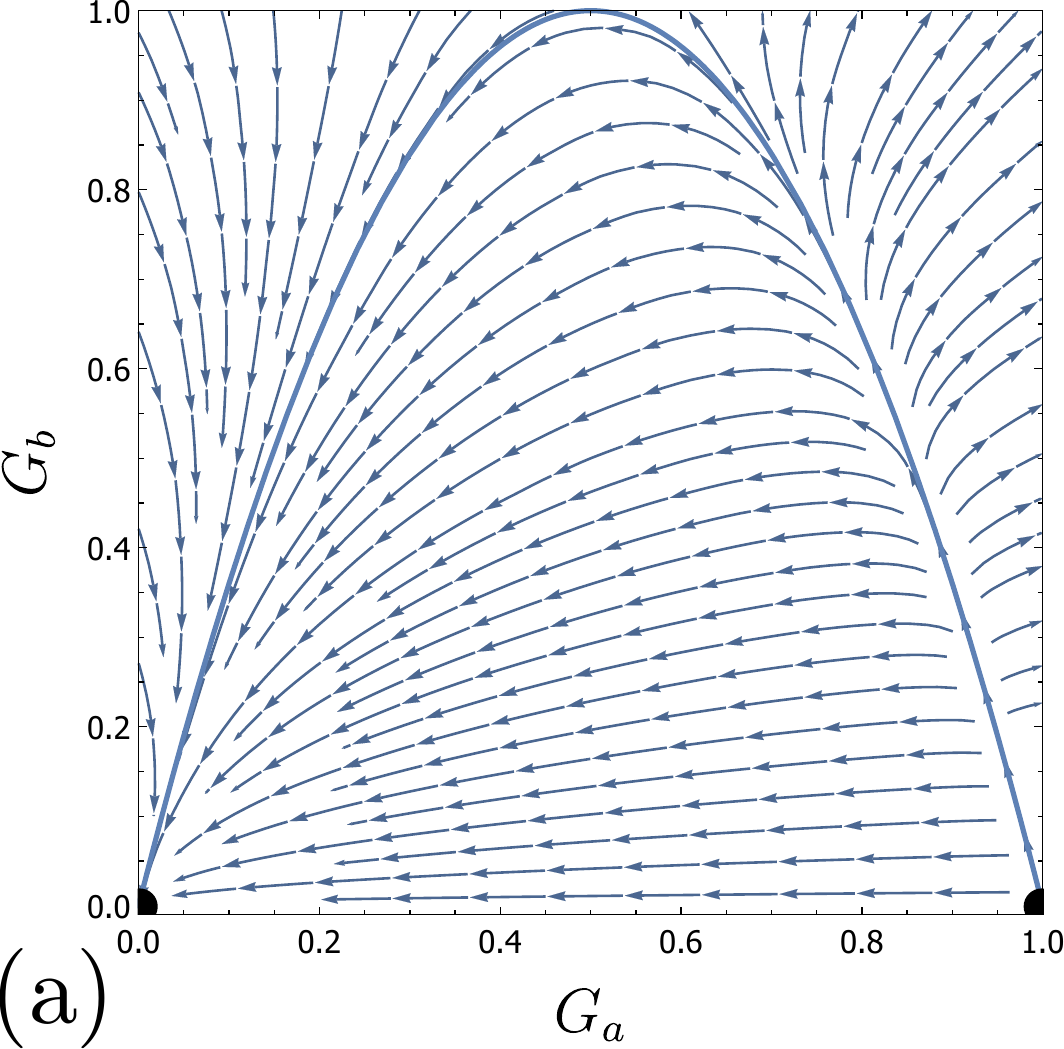} }
{\includegraphics[width=0.32\linewidth]{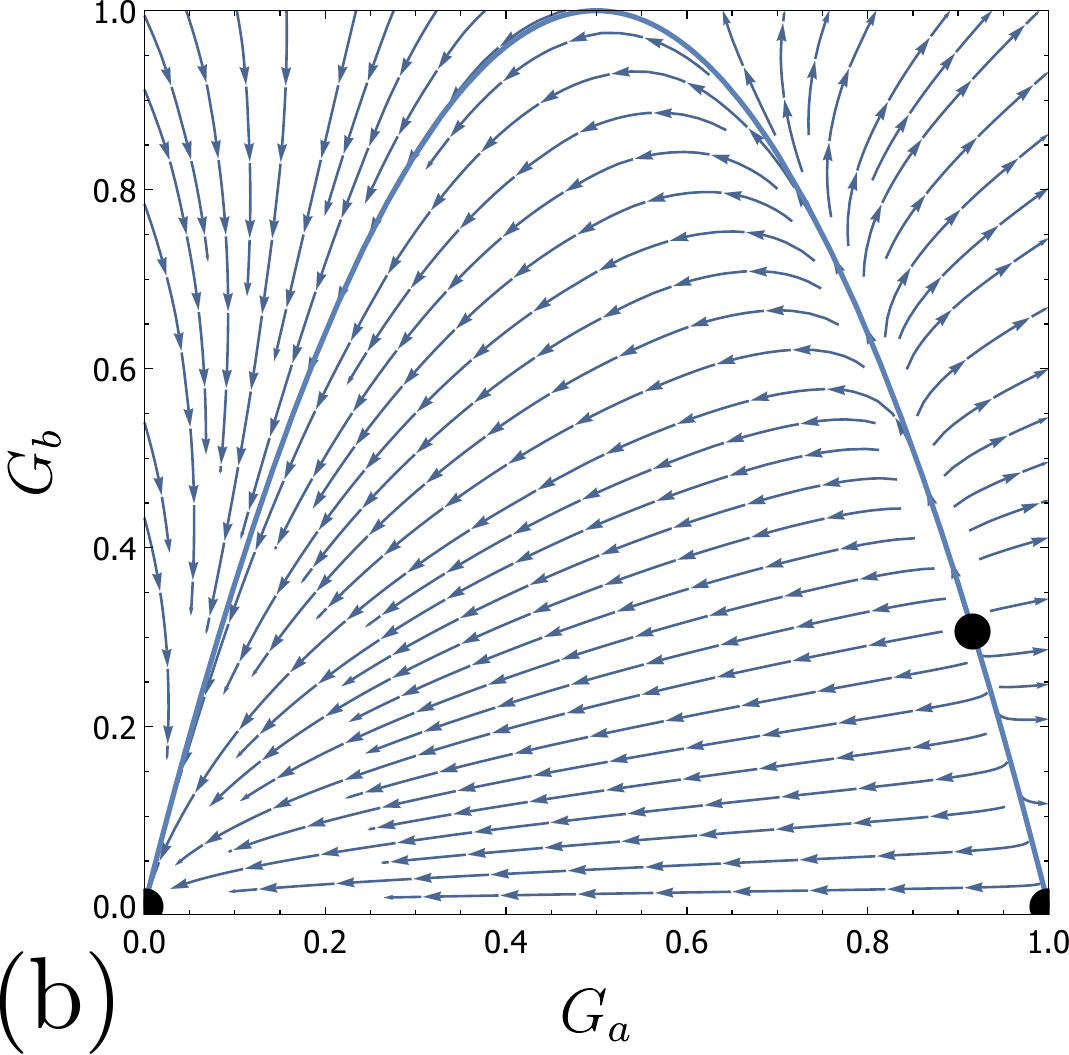} }
{\includegraphics[width=0.32\linewidth]{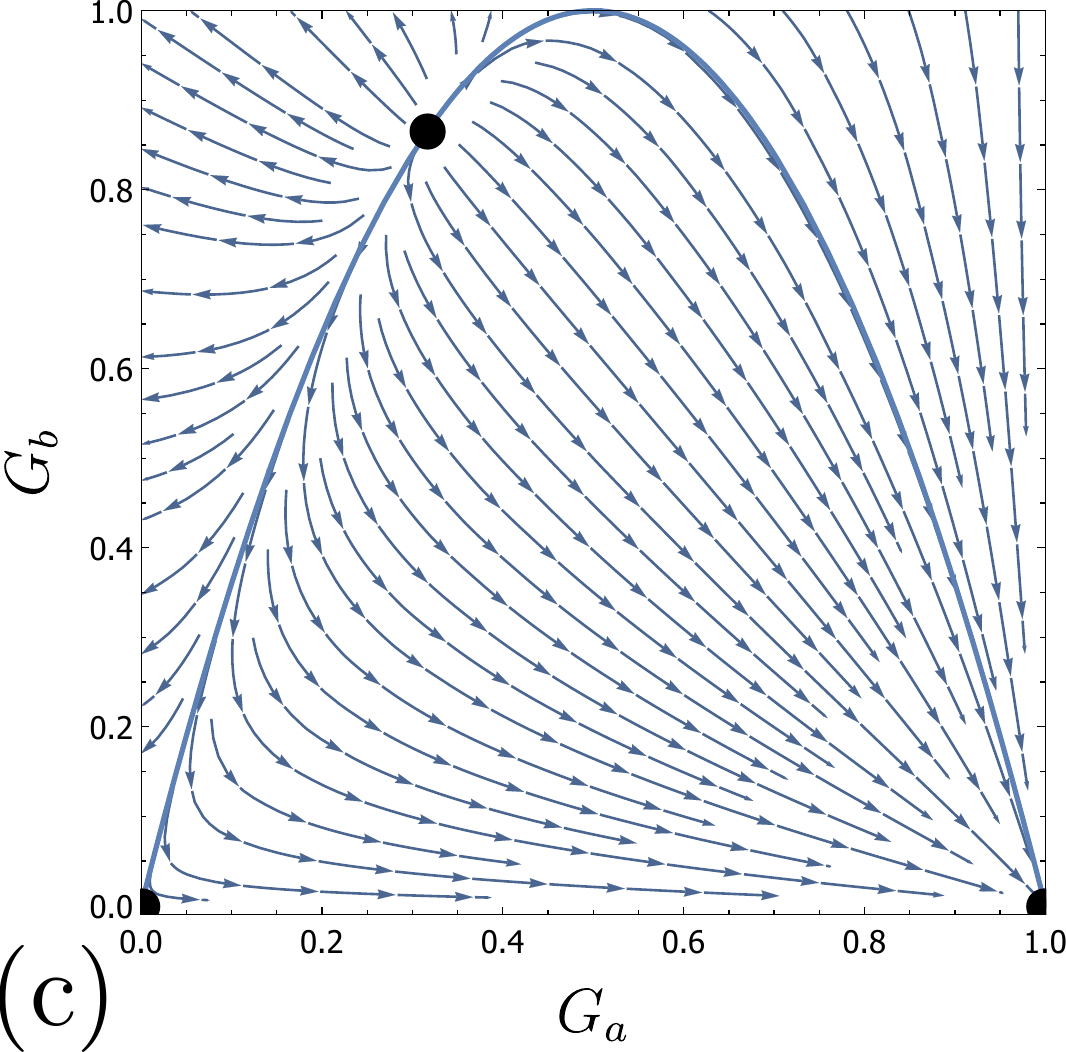} }
\caption{\label{fig:RGflowsEquil}
RG flows of conductances in equilibrium regime ($g=0.1$, $v/c=1.5$): (a)  full reflection of phonons, $\rho=1$, with electron-phonon interaction strength $\alpha=0.1$, (b)  ideal transmission of phonons, $\tau=1$, with electron-phonon interaction strength $\alpha=0.1$, (c)  ideal transmission of phonons, $\tau=1$, with electron-phonon interaction strength $\alpha=0.4$. The blue line delimits the area  of the physically available conductances.
}
\end{figure*}

First, we consider equilibrium limit $V_{a,b} \rightarrow 0$ and $\omega_D \rightarrow \infty$. The RG equations exhibit three fixed points:  point $N$ at $G_a = G_b = 0$ (complete junction breaking), point $A$ at $G_a = 1$, $G_b = 0$ (ideal transport in the main wire and the absence of the tunneling into the tip), and point $M$ at non-universal values of conductances given by
\aleq{
    G^{(M)}_a &=\frac{g+g_3-\bar\alpha (1+\xi \tau)}{g+2g_3-\bar\alpha},\\
    G_b^{(M)} &=1-\left(\frac{g-\bar\alpha(1+2\xi \tau)}{g+2g_3-\bar\alpha} \right)^2.
}
In presence of the  impurity non-transparent for phonons,  $\tau=0$, RG flows are similar to those without phonons (Fig.~\ref{fig:RGflowsEquil}(a)) with a modification of  scaling exponents as discussed below. When the phonons pass the impurity ($\tau>0$) the saddle fixed point $M$ appears (Fig.~\ref{fig:RGflowsEquil}(b)). 
This $M$ point exists in the first order in coupling constants even in the absence of interaction in the tip, $g_{3}=0$, contrary to the previous case of the pure local electron-electron interaction~\cite{Aristov2017}.

If the electron-phonon coupling is strong enough, it is possible for $M$ point to first move to the top of the parabola of allowed conductances and further pass to the left side of the RG diagram. This is accompanied by the ``metal-insulator''  transition~\citep{Galda2011} when  the stability of the fixed points $N$ and $A$ is interchanged  (see Fig.~\ref{fig:RGflowsEquil}~(c)). The marginal situation with the $M$ point located exactly at $G_a=1/2$ and $G_b=1$ corresponds to the existence of the whole line of fixed points at $G_b=0$.
This fixed line does not exist for any $g, g_{3} >0$ in the absence of the electron-phonon interaction. 

Non-equilibrium regimes of our system show variety of RG flows. There are three energy scales related with two bias voltages and one phonon energy scale. Depending on the running energy scale, $\epsilon$, different terms in the RG equations may contribute the behavior of conductances. The main difference from the previous non-equilibrium study of the Y junction~\cite{Aristov2017} is the presence of the Debye scale:  for the running energy $\epsilon$  below $\omega_D$, phonons start to contribute to the effective interaction. 

\begin{figure}
{\includegraphics[width=0.8\columnwidth]{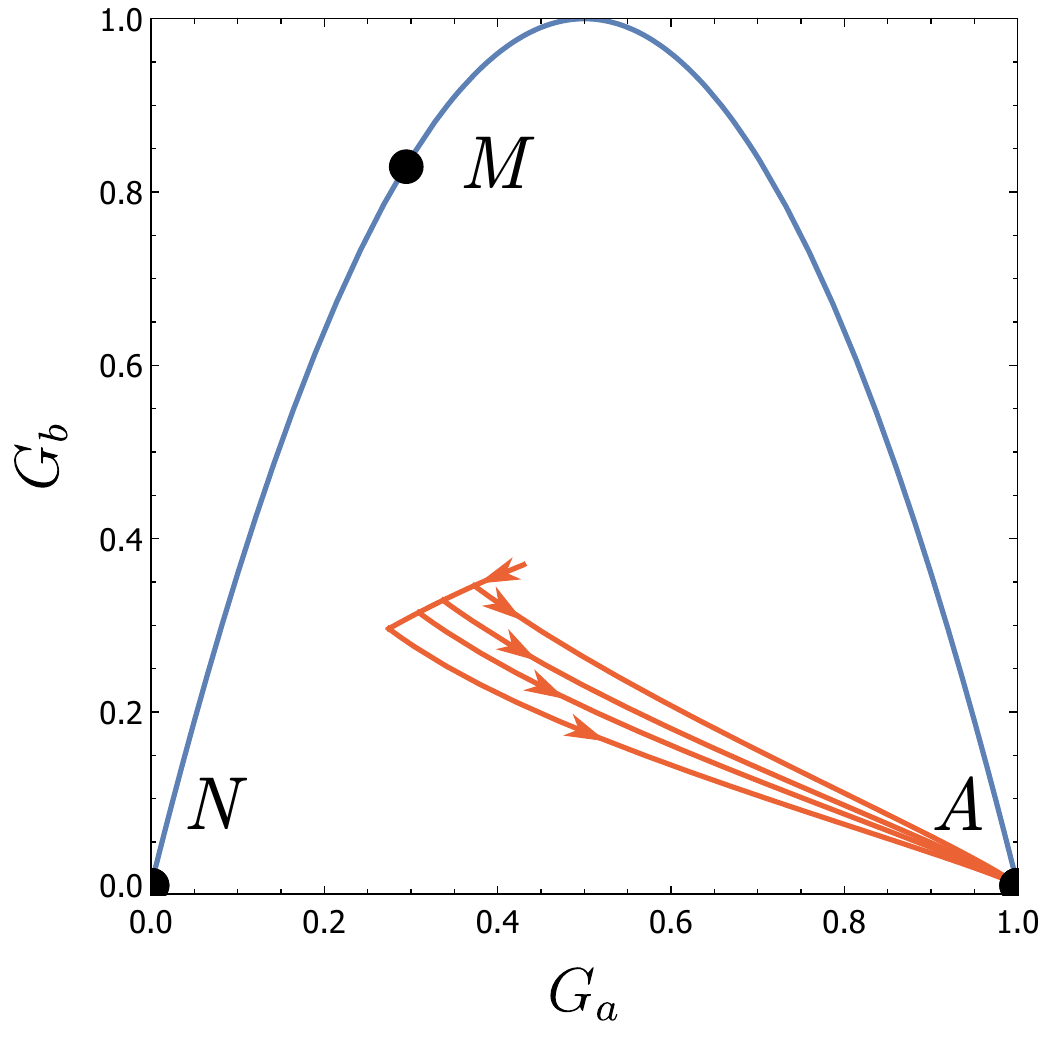} }
\caption{\label{fig:NonEqDebye}
Conductances RG flows in non-equilibrium regime in limit of zero temperature and voltages but finite Debye energy $\omega_D/\omega_0= 0.2, 0.07, 0.03, 0.01$ (from top to bottom) for phonons ideal transmission case, $\tau=1$, with interaction constants $\alpha=0.4$ and $g=0.1$, $v/c=1.5$ and the bare conductances $G_a(\omega_0)=0.43$, $G_b(\omega_0)=0.37$. 
}
\end{figure}

The importance of the intermediate scale,   $\omega_D$, (we assume that $\omega_D <\omega_0$ here and below) is perhaps best illustrated by the change in the position of $M$ point, which is not universal and is determined by the values of the coupling constants.
Indeed, at high energies (temperatures or voltages greater than the Debye frequency $ \omega_D $) there are no phonon contributions to RG, and the position of the $ M $ point is determined exclusively by the constants $ g $ and $ g_3 $. However, as the energy decreases below $\omega_{D}$,  the phonon corrections shift the $ M $ point or even lead to its appearance in case $ g_3 = 0$. This situtation results in the non-monotonic RG flow depicted in Fig.~\ref{fig:NonEqDebye}, and, potentially, to a change in its direction for  strong enough  electron-phonon interaction.

The effects of finite voltages were discussed previously in the literature and remain qualitatively the same. As before, the hierarchy of energy scales  is important. For example, in the regime  $V_a < \epsilon < \mu_2 \leq \mu_1 < \omega_{D}$ 
the RG flows for conductance $G_b$ terminate. Fixed points in this case form the parabola curve.
 The direction of flows is defined by the ``metallic''  or ``insulator''  character of the main wire.

\subsection{Fixed point analysis}\label{sec:fp1}
Let us first discuss the N fixed point.  Linearizing Eqs.~\eqref{RGeq1g} and introducing $G_c=G_a-\frac{1}{4}G_b$, we arrive at  the set of    RG equations
\aleq{\label{linearized_N}
\frac{dG_c}{d\Lambda}&=-2\Big(g-\bar\alpha(1+2\xi \tau)\theta_D(\epsilon)\Big)G_c\theta_a(\epsilon) \,,\\
    \frac{dG_b}{d\Lambda}&=-\frac{1}{2}\left(g+g_3-\bar\alpha(1+\xi \tau)\theta_D(\epsilon)\right)
G_b\theta_+(\epsilon) \,.
}
The renormalization occurs in several steps with different beta-functions at each step. In addition to three different regimes discussed in \cite{Aristov2017} the  new scale $\omega_D$ gives rise to further possible behavior of conductances. 

What concerns the behavior of $G_c$,  one can easily obtain the solution for $V_a  < \omega_{D}$ in the similar way as it was done in \cite{Aristov2017}. The renormalization of this quantity stops at $\epsilon <V_a$, so that

 \begin{equation}
\begin{aligned}
G_c(0) & =
G_c(\omega_0)\left(\frac{V_a}{\omega_0} \right)^{ 2g}  \,, \quad  \omega_{D}  <V_a \,, \\
& = G_c(\omega_0)\left(\frac{\omega_D}{\omega_0} \right)^{\kappa_{N,1}}\left(\frac{V_a}{\omega_0} 
 \right)^{ \gamma_{N,1}}  \,, \quad V_a  < \omega_{D} \,, 
\end{aligned}
\label{G_c_eq}
\end{equation}
with  scaling exponents 
$\gamma_{N,1}=2g-2\bar\alpha(1+2\xi \tau)$ and $\kappa_{N,1}= 2g-\gamma_{N,1}$.

The behavior of $G_b$ is more involved, depending on the relation between $\omega_D$ and $V_{b+}=\mu_{2}$.
Solving the corresponding equation for $G_b$ in case of $\omega_D>\mu_{2}>|\mu_{1}|$ one gets
\begin{equation}\label{G_b_eq}
    \frac{G_b(0)}{G_b(\omega_0)}=\left(\frac{\omega_D}{\omega_0} \right)^{\kappa_{N,2}}\left(\frac{\sqrt{\mu_{1}\mu_{2}}}{\omega_0} \right)^{\gamma_{N,2}}\,,
\end{equation}
with $\gamma_{N,2}=g+g_3-\bar\alpha(1+\xi \tau)$ and $\kappa_{N,2}=g+g_3-\gamma_{N,2}$. In the opposite case $\mu_{2}> \omega_D>|\mu_{1}|$ we obtain
\begin{equation}
    \frac{G_b(0)}{G_b(\omega_0)}=\left(\frac{\mu_{2}}{\omega_D} \right)^{-\frac{\kappa_{N,2}}{2}}\left(\frac{\sqrt{\mu_{1}\mu_{2}}}{\omega_0} \right)^{\left.\gamma_{N,2}\right|_{\alpha=0}}\,.
\end{equation}

Let us now describe the infrared character of this fixed point in equilibrium. For $ \gamma_{N,1}> 0 $ and $ \gamma_{N,2}> 0 $, the fixed point $ N $ is attractive, and the instability with respect to the formation of a charge density wave (CDW) renormalizes Friedel oscillations at the junction until all three wires are completely separated \cite{Aristov2010}. It is the situation when the $ M $ point is on the right hand side of the parabola of allowed conductances. Moreover, renormalization of the tunneling density of states leads to the vanishing of the conductance of the tunneling probe at $ \mu_1 = \mu_3=0 $ or $ \mu_2 = \mu_3=0 $.

For $ \gamma_{N,1} \gamma_{N,2}<0 $, the fixed point $ N $ is a saddle point. The point $ M $ moves to the left hand side of the RG diagram, but has not yet merged with the $ N $ point. Because we always have $ \gamma_{N,1} <0<\gamma_{N,2} $, the zero-bias anomaly still suppresses the conductance in the third wire, however $ G_a $ starts to grow below a certain scale. For $ \gamma_{N,1} <0 $ and $ \gamma_{N,2} <0 $ the point $N$ becomes unstable, because $ M $ passed through the $ N $ point at $ \gamma_{N,2} = 0 $.

Next we analyze the nonequilibrium scaling near the $A$ fixed point, $G_a=1$, $G_b=0$. We introduce the small displacement $\tilde{G}_a=1-G_a$ and the combination $\tilde{G}_c=\tilde{G}_a-\frac{1}{4}G_b$ and obtain
\aleq{
\frac{d\tilde{G}_c}{d\Lambda}&=2\Big(g-\bar\alpha(b_
\tau+b_\rho)\theta_D(\varepsilon) \Big)\tilde{G}_c\theta_a(\varepsilon)\,,\\
\frac{dG_b}{d\Lambda}&=-\frac{1}{2}\left(g_3+\bar\alpha b_\tau\theta_D(\epsilon)\right)
G_b\theta_+(\epsilon) \,.
}

This set of equations has the same structure as \eqref{linearized_N}, thus one can easily obtain renormalized $G_c$ and $G_b$ by simply replacing all scaling exponents to the following corresponding quantities:  $\gamma_{A,1}=-\gamma_{N,1}$, $\kappa_{A,1}=-\kappa_{N,1}$, $\kappa_{A,2}=-\bar\alpha\xi\tau$, and $\gamma_{A,2}=g_3-\kappa_{A,2}$. Introduced coefficients obey several symmetry relations such as $\gamma_{N,1}=2(\gamma_{N,2}-\gamma_{A,2})$ and $\kappa_{N,1}=2(\kappa_{A,2}+\kappa_{N,2})$.

One can notice that the presence of the  repulsive interaction $ g_3 $ in the tunneling tip affects the renormalization in the same way as the phonon attraction in the main wire. Thus, in order to obtain a nontrivial $ M $ point in the lowest order RG consideration, the interaction in the tip is not required at all and phonon exchange through the junction in the main wire effectively induces competition between the instability with respect to the formation of a charge density wave in the main wire and the renormalization of the tunneling density of states. Therefore, the $ A $ point in the presence of the electron-phonon interaction is either a saddle point or attractive (when the electron-phonon interaction is strong enough).

\section{RPA-type summation to infinite order in the interaction}\label{ladder_section}

So far we have obtained the beta functions for the conductances in the Y junction of quantum wires in the first order of perturbation theory with respect to the coupling constant of the electron-phonon interaction $ \alpha $~\eqref{RGeq1g}. The solution of the resulting RG equations is equivalent to summing the leading sequence of logarithms of the form $ \alpha^n \ln^n (\omega_0 / \epsilon)$.
We now turn to the investigation of higher-order corrections describing  relevant scale-dependent contributions to conductances. Our goal   is to include strong-coupling screening effects to the RG equations in framework of ``RPA-type approximation'',  as proposed in \cite{Aristov2014} for the case of the   short-range interactions out of equilibrium. The result of this procedure allows us to take into account the sub-leading logarithmic contributions from higher orders of perturbation theory. 

The RPA-type approximation involves dressing the local bare interaction with polarization fermionic loops. In systems with translational invariance, such a series of diagrams is reduced to geometric series and is easily summed up. However, in the scattering states formalism employed here the momentum is not conserved, and the summation of this RPA-like sequence is rather nontrivial.  In presence of interactions with  non-local character and retardation effects, Eq.~\eqref{eq:greenphon}, the summation procedure becomes even more involved  due to the complicated form of the bare bosonic propagator.

The explicit form of the integral equation which describes the summation of the RPA sequence of the diagrams  is as follows
\aleq{\label{WH_0}
    \widehat{\mathbf{L}}^R(x|y)=\widehat{\mathbf{L}}^{(0)}(x|y)- \int  dz_{1}dz_2 \;\widehat{\mathbf{L}}^{(0)}(x|z_1)\;\\
    \times
\begin{bmatrix}
\Pi(-z_1 |-z_2)\mathbf{1}  & 0 \\
    \Pi(z_1|-z_2)\mathbf{Y}   & \Pi(z_1 |z_2)\mathbf{1} \\
\end{bmatrix}
\widehat{\mathbf{L}}^R(z_2|y)\,,
}
where $Y_{ij}=|S_{ij}|^2$, and $\widehat{\mathbf{L}}^{(0)}(x|y)$ is defined in \eqref{L0_bare}. We dropped all unimportant labels here and introduced a dynamical factor $\Pi(x | z)=(2\pi v^2)^{-1}[v\delta(x~-~z)+i\omega\theta(x-z)e^{i\frac{\omega}{v}(x-z)}]$. The kernel in \eqref{WH_0} corresponds to the fermionic loop calculated with  Green's functions \eqref{GF_def} (see \cite{Aristov2014} for details).

This section is devoted to exact solution of the integral equation \eqref{WH_0}, which allows us to analyze the strong-coupling limit. The result of this rather cumbersome calculation is given by Eqs.~\eqref{C_solution} and \eqref{L_solution} below. 

As a first step of our calculation we set $Y_{ij}=0$ in  \eqref{WH_0}, thus discarding all contributions 
containing  matrix elements of the $S$ matrix. The remaining sum defines an auxiliary 
interaction $ \mathbf{C} $, which incorporates  strong-coupling effects taking place far away from the junction. In contrast to the previously studied cases of short-ranged interactions, \cite{Aristov2014} the quantity $ \mathbf{C} $ cannot be fully attributed to the ``bulk" of the main wire because of the presence of the boundary terms in \eqref{WH_0} proportional to the $B$ matrix. These terms describe 
phonon scattering processes in the vicinity of the junction and lead to the off-diagonal structure of $\mathbf{C}$ in wire space. In terms of this new propagator, the full dressed interaction $ \widehat{\mathbf{L}}^R $ can be represented as
\aleq{\label{WH_1}
    \widehat{\mathbf{L}}^R(x|y)=\widehat{\mathbf{C}}(x|y)- \int  dz_{1}dz_2 \;\widehat{\mathbf{C}}(x|z_1)\;\\
    \times
\begin{bmatrix}
0  & 0 \\
    \Pi(z_1|-z_2) \mathbf{Y}  & 0 \\
\end{bmatrix}
\widehat{\mathbf{L}}^R(z_2|y)\,.
}

 Since the integral kernel for $\widehat{\mathbf{C}}$ is diagonal in wire indices and the bare line itself does not connect the main wire and the tip, then the equation is essentially split.   
It allows us to focus  on the main wire in the analysis of $\widehat{\mathbf{C}}$ and assume all matrices   reduced to their 2$\times$2 sub-blocks (and set $\lambda_{1,2} = \lambda$). Further we note  that $ \widehat{\mathbf{C}}$ does not depend on the chiral structure, and we can write $\widehat{\mathbf{C}}=\mathbf{C}(\widehat{\tau}^0+\widehat{\tau}^1)$. It is helpful to introduce the  symmetric combination
\aleq{
    \Pi^{s}(x|y)&=\Pi(x|y)+\Pi(-x|-y)\\
    &=\frac{1}{2\pi v^2}\left(2v\delta(x-y)+i\omega\; e^{i\frac{\omega}{v}|x-y|} \right) \,,
}
in terms of which the integral equation acquires the form
\aleq{\label{WH_2}
&\mathbf{C}(x|y)=\mathbf{L}^{(0)}(x|y)
- \lambda \int   dz  \Pi^{s}(x|z)\mathbf{C}(z|y)\\
&-i\omega \zeta \hspace{-0.2em} \int \hspace{-0.2em} dz_{1,2} \Pi^s(z_1|z_2)\big(e^{i\frac{\omega}{c}|x-z_1|}\mathbf{1}+e^{i\frac{\omega}{c}(x+z_1)} \mathbf{B}\big)\mathbf{C}(z_2|y).
}
with $dz_{1,2}=dz_1dz_2$. Thus, we have reduced the initial problem to the set of two  integral equations \eqref{WH_1} and \eqref{WH_2} with a transparent physical meaning: Eq.~\eqref{WH_2} describes screening processes  in the bulk of the main wire, and Eq.~\eqref{WH_1} includes  scattering events at the junction encoded in the $\mathbf{Y}$ matrix.

Before proceeding   further, let us highlight two main distinctive features of these integral equations as compared to purely local interactions ($\alpha=0$) discussed in \cite{Aristov2014}. First of all, the bare interaction $\mathbf{L}_0$ depends on coordinates and frequencies in a non-trivial way capturing retardation effects due to scattering on phonons. This fact complicates the integral equation \eqref{WH_2} in the ``kinematic'' sense.

\begin{figure}[h!]
\includegraphics[width=0.8\columnwidth]{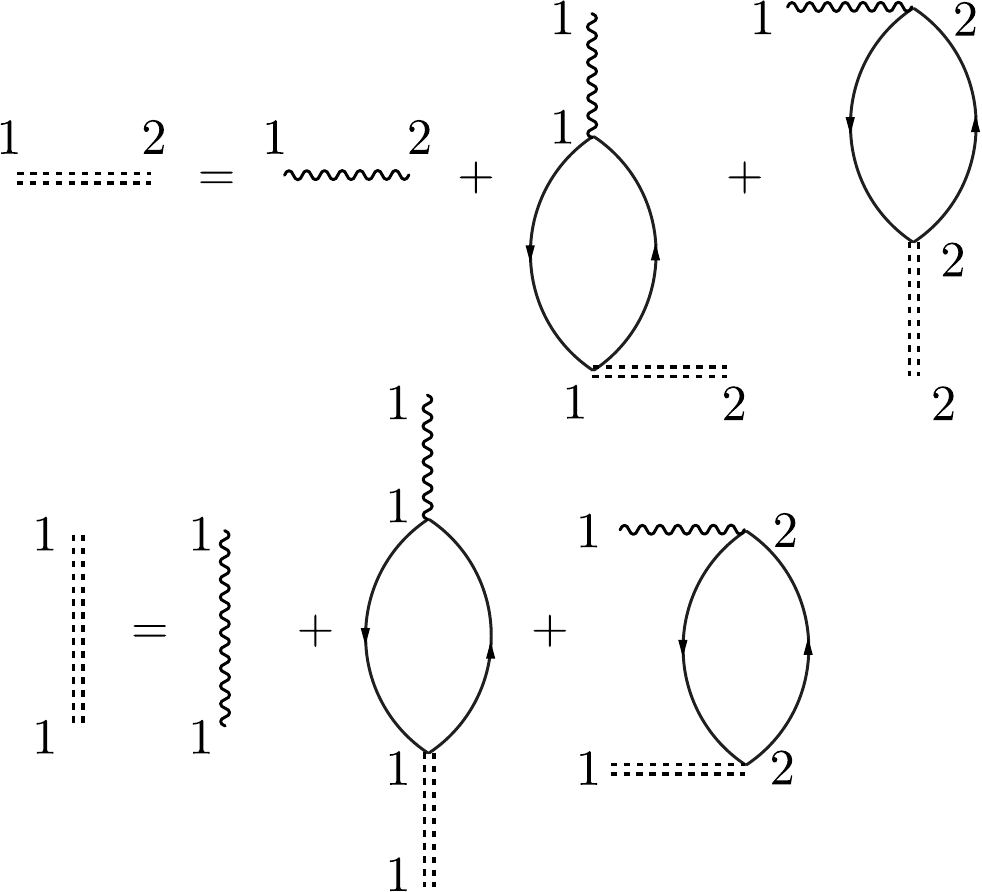}
\caption{{Integral equations for the auxiliary quantity $\mathbf{C}$ are shown in the``main wire'' space: indices corresponds to wire labels. Vertical lines 1--1, 2--2 correspond to the``local'' processes in the same wire. Horizontal lines 1--2 are proportional to the non-diagonal elements of $\mathbf{B}$ matrix~\eqref{eq:Bmat} and correspond to the``non-locality'' of the phonon-mediated interaction. Fermionic loops contain only diagonal components with $\mathbf{Y}=0$.}
}
\label{Fig:Eq_C}
\end{figure}

Second, the bare interaction $\mathbf{L}_0$ is now non-diagonal in wire index due to the propagation of phonons through the junction with transmission and reflection amplitudes, $B_{lm}$  (generally independent of electrons amplitudes, $S_{lm}$). This property has important consequences as discussed below. The Fig.~\ref{Fig:Eq_C} depicts concrete non-diagonal processes taken into account in \eqref{WH_2}. We employed there the following diagrammatic rules: all vertical lines (loops) correspond to diagonal matrix elements of propagators (polarization operators) with the same wire index, and horizontal lines represent off-diagonal contributions due to non-zero $\tau$ in \eqref{eq:Bmat}. The integration over positions is assumed for each vertex.

\subsection{Reduction to the linear differential equation}

Despite the apparent complexity of \eqref{WH_2}, it   can be reduced to the linear ordinary differential equation by repeated differentiation with respect to $ x $.   
As an intermediate step we introduce the following integral quantities
\aleq{\label{integrals}
    &\mathbf{I}_{\beta}= \int  dz \; e^{i\frac{\omega}{\beta}|x-z|}\mathbf{C}(z|y),\quad \mathbf{J}_{\beta}=\int  dz \; e^{i\frac{\omega}{\beta}(x+z)} \mathbf{B} \mathbf{C}(z|y)   \,, \\
     &\mathbf{Q}=\int  dz_{1,2}  e^{i\frac{\omega}{v}|z_1-z_2|}\left(e^{i\frac{\omega}{c}|x-z_1|}\mathbf{1}+e^{i\frac{\omega}{c}(x+z_1)}\mathbf{B}\right)\mathbf{C}(z_2|y)  \,,
}
with the omitted explicit coordinate dependence in   $\mathbf{I},\mathbf{J}$ and $\mathbf{Q}$. One can     verify the following relations for the derivatives
\aleq{\label{eq:relations_int}
   \partial_x^2 \mathbf{I}_{\beta}=-\frac{\omega^2}{\beta^2}\mathbf{I}_{\beta}+&\frac{2i\omega}{\beta}\mathbf{C}\,, \quad
     \partial_x^2 \mathbf{J}_{\beta}=-\frac{\omega^2}{\beta^2}\mathbf{J}_\beta   \,, \\
 \partial_x^2 \mathbf{Q}&=-\frac{\omega^2}{c^2}\mathbf{Q}+\frac{2i\omega}{c}\mathbf{I}_v \,.
}
Using these relations, we can  express the integral equation \eqref{WH_2} in the compact way as
\begin{equation}\label{WH_equation_compact1}
2\pi v^2\tilde{d}^2\mathbf{C}= 2\pi v^2 \mathbf{L}^0 -i\omega\lambda \mathbf{I}_v -2i\omega v \zeta \left(\mathbf{I}_c +\mathbf{J}_c \right) +\omega^2\zeta \mathbf{Q}  \,,
\end{equation}
where $\tilde{d}^2=1+2\tilde{g}$ and $\tilde{g}=g-\alpha/2$.

We notice that the kernel in Eq.\eqref{WH_equation_compact1} has a jump in its derivative at $x = z$, which we use by twice
differentiating it with respect to $x$. We thus arrive at a second-order integro-differential equation
\aleq{\label{WH_equation_compact2}
2\pi v^2  \mathbf{C}^{(2)}= 2\pi v^2 \left(\mathbf{L}^0\right)^{(2)} +4\omega^2 \gamma \mathbf{C}
+ 2i\omega^3\frac{\gamma}{v} \mathbf{I}_v\\
-\omega^2c^{-2}\left[
-2i\omega v\zeta \left(\mathbf{I}_c+\mathbf{J}_c \right) +
\omega^2\zeta  \mathbf{Q} \right]\,,
}
where, for simplicity, we introduced $\gamma=\lambda/2v+\zeta v/c$. The label $(n)$ in $ \mathbf{C}^{(n)}$ stands for the $n$-th derivative with respect to $x$. 
The combination in the square brackets in \eqref{WH_equation_compact2} is eliminated  by using 
Eq.~\eqref{WH_equation_compact1} and we obtain 
 \begin{equation}
\begin{aligned}
& 2\pi v^2 \tilde{d}^2 \mathbf{C}^{(2)} =  2\pi v^2 \left(\mathbf{L}^0\right)^{(2)}+2\pi\omega^2 v^2 \mathbf{L}^0 c^{-2}   \\
 & +\omega^2\left(4\gamma - \chi c^{-2}\right)\mathbf{C}
+i\omega^3 \left(2\gamma v^{-1}-\lambda c^{-2}\right)\mathbf{I}_v   \,. \\
\end{aligned}
\label{WH_equation_compact2a}
\end{equation}
The last equation still contains $\mathbf{I}_v$. 
Differentiating it again twice,   we find 
\aleq{
&2\pi v^2 \mathbf{C}^{(4)} =  2\pi v^2 \left(\mathbf{L}^0\right)^{(4)}+2\pi\omega^2 v^2\left(\mathbf{L}^0\right)^{(2)}c^{-2}
\\
&+\omega^2\left(4\gamma - \chi  c^{-2}\right)\mathbf{C}^{(2)} -2\omega^4/v \left(2\gamma v^{-1}-\lambda c^{-2}\right)\mathbf{C}\\
&-i\omega^4 v^{-2} \left(2\gamma v^{-1}-\lambda c^{-2}\right)\mathbf{I}_v \,.
}
We can eliminate $\mathbf{I}_v$ here  by expressing it from  the \eqref{WH_equation_compact2a}. Then we finally arrive at a linear inhomogeneous ordinary differential equation of  fourth order
\aleq{\label{diff_eq_C}
\tilde{d}^2 \mathbf{C}^{(4)}+\omega^2\left(\frac{2g}{c^2}+\frac{1}{c^2}+\frac{1}{v^2}\right)\mathbf{C}^{(2)}+\frac{\omega^4}{v^2c^2}\mathbf{C}\\
=  \left(\mathbf{L}^0\right)^{(4)}+\omega^2\left(\frac{1}{v^2}+\frac{1}{c^2}\right) \left(\mathbf{L}^0\right)^{(2)}+\frac{\omega^4}{v^2c^2}\mathbf{L}^{0}\,.
}

Remarkably, this equation can be represented in terms of differential operators $D_{\rm{v}}=\partial^2_x+\omega^2/\rm{v}^2$ in the very compact form
\begin{equation}\label{diff_eq_C_operator}
D_{v_+}D_{v_-}\mathbf{C}(x|y)=\tilde{d}^{-2}D_{v}D_{c}\mathbf{L}^{0}(x|y)\,.
\end{equation}

A key feature of the obtained differential equation is that its homogeneous solution   can be expressed as the sum of   exponents $ e^{\pm i {\omega x}/{v_\pm}} $ with characteristic velocities given by
\aleq{
\label{speed}
v^2_\pm=\frac{1}{2}\Big( d^2 v^2+c^2 \pm \sqrt{\big(d^2v^2-c^2\big)^2+4\alpha v^2 c^2 }\Big) \,,
}
where $ d^2 = 1 + 2g $. The obtained characteristic velocities \eqref{speed} are nothing else but two hybridized polaron modes that arise in the non-perturbative bosonization treatment of the problem   \cite{Galda2011}.

In passing, the structure of the electron-electron interaction~\eqref{eq:elel0} in chirality indices implies that  $\mathrm{g}_4=\mathrm{g}_2 =g$ in $\mathrm{g}$-ology vocabulary of one dimensional studies. 
The above renormalization factor of the Fermi velocity has a well known form $ d^2 = (1 + \mathrm{g}_4) ^ 2-\mathrm{g}_2^2$, and for $ \mathrm{g}_4 = 0 $ one obtains $ d^2 = 1-g^2 $ in full agreement with \cite{Aristov2011}.

We now turn to the analysis of the non-homogeneous solution of the equation~\eqref{diff_eq_C}. First, we find the explicit expression for its right hand side (see Appendix~\ref{App_operator_relations} for further details):
\aleq{
D_{v}D_{c}&\mathbf{L}^{0}(x|y)=\lambda \delta^{(4)}(x-y)\mathbf{1}
+\omega^2\left(\frac{\lambda}{v^2}+\frac{\lambda}{c^2}-\frac{2\zeta}{c}\right)
\\&\times\delta^{(2)}(x-y)\mathbf{1}+\omega^4\left(\frac{\lambda}{c^2 v^2}-\frac{2 \zeta}{c v^2} \right)\delta(x-y)\mathbf{1}  \,.
}
We seek the solution $\tilde{\mathbf{C}}(x|y)$ of the inhomogeneous differential equation \eqref{diff_eq_C_operator} in the   form
\begin{equation}
\tilde{\mathbf{C}}(x|y)=\kappa_0\delta(x-y)\mathbf{1}+\frac{i\pi\omega}{\tilde{d}^2}
    \sum\limits_{\sigma=\pm}\kappa_\sigma e^{i\frac{\omega}{v_\sigma}|x-y|}\mathbf{1} \,.
\end{equation}
After some algebra we obtain coefficients $\kappa_{i}$ explicitly as
\begin{equation}
\kappa_0=\frac{2\pi v \tilde{g}}{\tilde{d}^2}, \quad \kappa_\pm=\frac{v^2_\mp}{c^2} \left(\frac{v^2-v^2_\pm}{v^2_\pm -v^2_\mp}\right) \frac{(gv^2_\pm -\tilde{g} c^2)}{ v v_\pm}  \,.
\end{equation}
Accordingly, the full solution for \eqref{diff_eq_C_operator} reads as
\aleq{ \label{C_full}
\mathbf{C}(x|y)=\kappa_0\delta(x-y) \mathbf{1}+\frac{i\pi\omega}{\tilde{d}^2}
    \sum\limits_{\sigma=\pm}\kappa_\sigma e^{i\frac{\omega}{v_\sigma}|x-y|}\mathbf{1} \\
    - \frac{i \pi \omega }{\tilde{d}^2}\sum\limits_{\sigma,g=\pm}\mathbf{A}_{\sigma g}(y)\;e^{g\; i\frac{\omega}{v_\sigma}x} \,, 
}
with yet unknown matrices $\mathbf{A}_{sg}$ which should be determined from the initial integral equation \eqref{WH_2}. The first two terms in \eqref{C_full} are translationally invariant and, thus, independent of the junction and should be associated with bulk effects. On the other hand, the last term has a factorized coordinate dependence and originates not only from the fermionic S matrix in Eq.~\eqref{GF_def} but also from the phonons' $\mathbf{B}$ matrix. 

Now we substitute the ansatz \eqref{C_full} into the initial   equation \eqref{WH_equation_compact1} and compare coefficients corresponding to different linearly-independent $x$-functions. This procedure is straightforward, albeit cumbersome, so we just present the resulting linear equations here. Further details are given in the Appendix \ref{reduction_A}.

\begin{figure*}[!t]
{\includegraphics[width=0.95\linewidth]{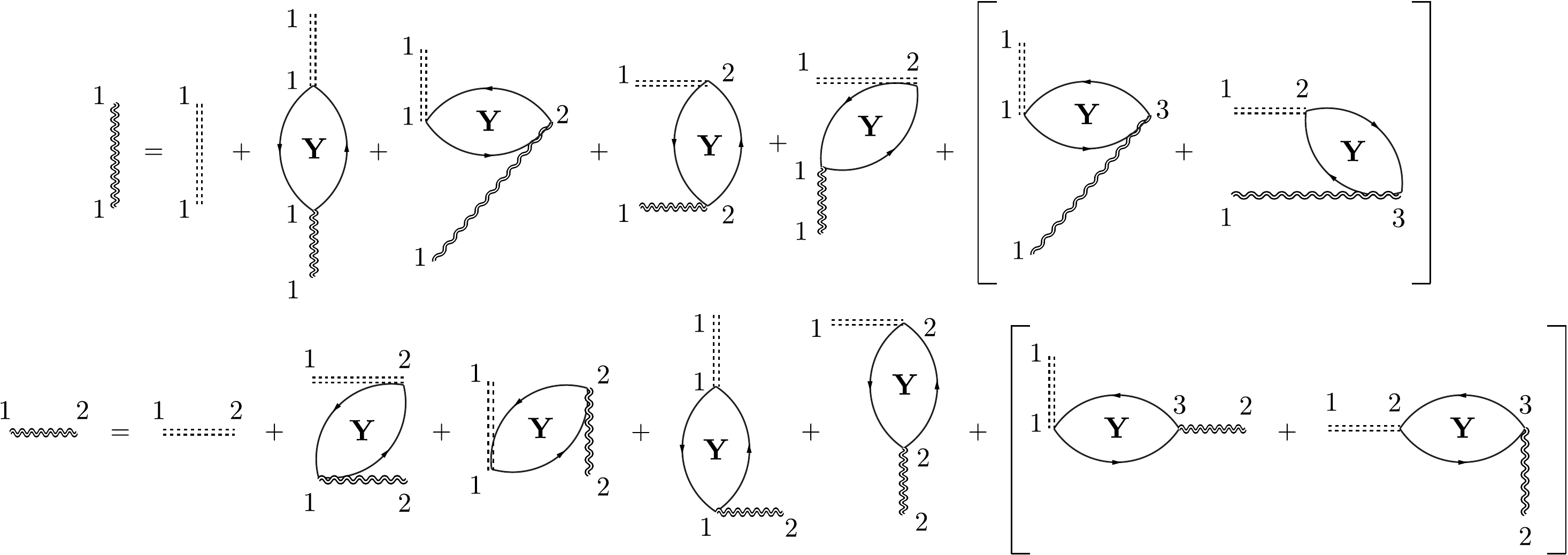} }
\caption{Integral equations for the fully dressed bosonic propagator $\mathbf{L}$ in the wire-space representation. The double-dashed line is the auxiliary quantity $\mathbf{C}$ given by the integral equation \eqref{WH_2}. Vertical lines $1-1$, $2-2$ correspond to the``local'' interaction in the same wire screened by all bulk effects. Horizontal lines $1-2$ are proportional to the non-diagonal elements of $\mathbf{B}$ matrix~\eqref{eq:Bmat}. In addition, horizontal and diagonal fermionic loops represent scattering processes of the junction. The label``Y'' stands for the contributions to the fermionic loop originated from matrix elements of $\mathbf{Y}$. Square brackets represent the processes involving tunneling into the third wire. Explicitly shown equations should be accompanied by one extra equation with a similar structure for the $1-3$ components of the propagator $\mathbf{L}$.
}\label{fig:eq_L}
\end{figure*}

We find that the  matrices $\mathbf{A}_{\sigma g}$ satisfy a following set of boundary conditions
\aleq{ \label{Full_Eq1}
    \sum\limits_{\sigma,g=\pm} \tilde{\chi}_1^{v,gv_\sigma}\mathbf{A}_{\sigma g} (y) &= \sum\limits_{\sigma=\pm}\kappa_\sigma \tilde{\chi}_{1}^{v,-v_\sigma}e^{i\frac{\omega}{v_\sigma}y} \mathbf{1}\,, \\
    \sum\limits_{\sigma,g=\pm} \tilde{\chi}_2^{v,gv_\sigma}\mathbf{A}_{\sigma g} (y) &=
    \sum\limits_{\sigma=\pm}\kappa_\sigma \tilde{\chi}_{2}^{vv_\sigma}e^{-i\frac{\omega}{v_\sigma}y}\mathbf{1}\,, \\
     \sum\limits_{\sigma,g=\pm} \phi_{g+}^{\sigma}\mathbf{A}_{\sigma g} (y) &=
    \sum\limits_{\sigma=\pm}\kappa_\sigma \phi_{++}^{\sigma} e^{-i\frac{\omega}{v_\sigma}y}\mathbf{1}\,, \\
     \sum\limits_{\sigma,g=\pm} \tilde{\mathcal{B}}_g^\sigma \mathbf{A}_{\sigma g} (y)  &=
     \sum\limits_{\sigma=\pm}\kappa_\sigma \tilde{\mathcal{B}}_-^\sigma e^{i\frac{\omega}{v_\sigma}y}\,,
}
with the following scalar coefficients 
\aleq{
\tilde{\chi}_1^{\beta_1\beta_2}&=\frac{ \beta_1 \beta_2}{ \beta_1-\beta_2},\quad \tilde{\chi}_2^{\beta_1\beta_2}=\tilde{\chi}_1^{\beta_1,-\beta_2} e^{i\omega L\left(\frac{1}{\beta_1}+\frac{1}{\beta_2} \right)}\,,\\
&\phi_{gs}^\sigma =\tilde{\chi}_2^{c,g v_\sigma}+s\frac{c^2}{ v^2}\frac{ \tilde{\chi}_1^{v,-sc}\tilde{\chi}_1^{v,g v_\sigma}\tilde{\chi}_2^{v,g v_\sigma}}{\tilde{\chi}_1^{v,-c}\tilde{\chi}_2^{v,-c}}\,.
}

A matrix  entering the last equation in \eqref{Full_Eq1} is defined  as
\begin{equation}
    \tilde{\mathcal{B}}_g^\sigma=\phi_{-g,+}^\sigma e^{gi\frac{\omega}{v_\sigma}L}\mathbf{1}-  \phi_{g,-}^\sigma e^{-gi\frac{\omega}{v_\sigma}L}\mathbf{B}\,.
\end{equation}

Despite the apparent progress, the linear system \eqref{Full_Eq1} still looks rather complicated. 
Further progress is achieved by diagonalizing all matrices in the wire space with  the unitary transformation
\begin{equation}
    \mathbf{U}=\frac{1}{\sqrt{2}}\begin{pmatrix}
    1 & 1\\
    1 & -1
    \end{pmatrix}, \quad \mathbf{U}\mathbf{B}\mathbf{U}=\begin{pmatrix}
    \rho+\tau & 0\\
    0 & \rho-\tau
    \end{pmatrix} \,. 
\end{equation}
After that the diagonal  $\mathbf{A}_{\sigma j}(y)$  can be  replaced  by its corresponding diagonal matrix element $A_{\sigma j}(y)$. Next simplification comes from the representation 
\begin{equation}
    A_{\sigma j}(y)=\sum \limits_{s,g=\pm} \kappa_s b^{sg}_{\sigma j} \;e^{g\;i\frac{\omega}{v_s}y}\,.
\end{equation}
Indeed, for the new variables, $b^{sg}_{\sigma j}$, the system \eqref{Full_Eq1} factorizes into four decoupled sectors parametrized by indices $s$ and $g$. The appearing sets of algebraic equations can be easily solved by means of computer algebra methods (for instance, in {\it Mathematica}). The explicit form of all coefficients $b^{sg}_{\sigma j}$ is presented in App.~\ref{b_solution}.

The final expression for $\mathbf{C}$ which  solves \eqref{WH_2} exactly for an {\it arbitrary} set of parameters is given by
\aleq{\label{C_solution}
    \mathbf{C}(x|y)=\kappa_0\delta(x-y) \mathbf{1}+\frac{i\pi\omega}{\tilde{d}^2}
    \sum\limits_{\sigma=\pm}\kappa_\sigma e^{i\frac{\omega}{v_\sigma}|x-y|}\mathbf{1}\\
    -\frac{i\pi \omega}{\tilde{d}^2}\sum\limits_{\sigma,j=\pm} e^{j\frac{i\omega x}{v_\sigma}}\sum\limits_{s,g=\pm}\kappa_s \mathbf{U}\mathbf{b}^{sg}_{\sigma j}\mathbf{U}\; e^{g\frac{i \omega y}{v_s}}\,,
}
where $\mathbf{b}^{sg}_{\sigma j}=\text{diag}\left[\left.b^{sg}_{\sigma j}\right|_{B\rightarrow \rho+\tau}, \;\left.b^{sg}_{\sigma j}\right|_{B\rightarrow \rho-\tau} \right]$, see App.~\ref{b_solution}.

Summarizing this subsection, we have solved the integral equation \eqref{WH_2} for the main wire. We stress that the resulting ``bulk'' propagator has a non-diagonal form \eqref{C_solution} which should be understood as a 2$\times$2 sub-block of the full 3$\times$3 matrix $\mathbf{C}$. The remaining diagonal matrix element corresponds to the tunneling tip and can be obtained from \eqref{C_solution} by setting $\alpha=0$ and replacing $g$ by $g_3$. 

\subsection{Full equation for $L$}
Now we have everything at hand to solve Eq.~\eqref{WH_1} exactly. The non-diagonal structure of $\mathbf{C}$ significantly complicates the set of scattering processes contributing to the fully dressed propagator $\mathbf{L}$ which is depicted in Fig.~\ref{fig:eq_L}. The diagrammatic rules here are slightly more complicated compared to Fig.~\ref{Fig:Eq_C} because we include 
interwire 
parts of polarization loops proportional to fermionic transmission and reflection amplitudes encoded in $\mathbf{Y}$. As before, we use vertically orientated objects (propagators and loops) to describe processes diagonal in wire indices, and horizontally oriented (or tilted) ones for off-diagonal contributions. The full Eq.\ \eqref{WH_1} mixes contributions from different wires, including a tunneling tip, which was previously decoupled in \eqref{WH_2}, therefore we also consider diagrams with tunneling processes through a third wire, see caption of Fig.~\ref{fig:eq_L} for additional details.

 The integral equation \eqref{WH_1}, however, has a separable kernel and we  can easily solve it by rewriting it as  
\begin{equation}
    \widehat{\mathbf{L}}^R(x|y)=\widehat{\mathbf{C}}(x|y)+\int dz_{1,2} \;\widehat{\mathbf{C}}(x|z_1) \widehat{\mathbf{\mathcal{Y}}}(z_1|z_2)
\widehat{\mathbf{C}}(z_2|y)\,,
\end{equation}
where we introduced the summation of all fermionic loop contributions proportional to $\mathbf{Y}$ with the propagator $C$ as a new kernel 
\aleq{\label{Y_int_eq}
    &\widehat{\mathbf{\mathcal{Y}}}(x|y)=-\frac{i \omega}{2 \pi v^2} e^{i\frac{\omega}{v}(x+y)} \;
\begin{bmatrix}
0  & 0 \\
    \mathbf{Y}   & 0 \\
\end{bmatrix}
\\
-& \frac{i \omega}{2 \pi v^2} e^{\frac{i\omega x}{v}}\;
\begin{bmatrix}
0  & 0 \\
    \mathbf{Y}   & 0 \\
\end{bmatrix}\int dz_{1,2} \;e^{\frac{i\omega z_1}{v}}\widehat{\mathbf{C}}(z_1|z_2) \widehat{\mathbf{\mathcal{Y}}}(z_2|y)\,.
}
The full solution to \eqref{Y_int_eq} has a form
\begin{equation}
    \widehat{\mathbf{\mathcal{Y}}}(x|y)=-\frac{i \omega}{2 \pi v^2} e^{i\frac{\omega }{v}(x+y)} \mathbf{Y}
   \left( \frac{1}{\mathbf{1}+\frac{i \omega}{2 \pi v^2}\mathbf{C}_{s}\mathbf{Y}}\right) \begin{bmatrix}
0  & 0 \\
    1   & 0 \\
    \end{bmatrix},
\end{equation}
where the integrated quantitiy $\mathbf{C}_{s}$ is 
\begin{equation} \label{C_simple}
   \mathbf{C}_{s}= \int dx dy \;\mathbf{C}(x|y)e^{i\frac{\omega }{v}(x+y)}\,,
\end{equation}
and label ``s'' standing for``simplified". We note that   the full propagator does not depend on chirality indices $ \widehat{\mathbf{L}}^R= \mathbf{L}^R(\widehat{\sigma}^0+\widehat{\sigma}^1)$, so  we can analyze $\mathbf{L}^R$.

As a result, the fully-dressed interaction propagator is obtained in the following form
\begin{equation}\label{L_solution}
    \mathbf{L}^R(x|y)=\mathbf{C}(x|y)-\frac{i \omega}{2 \pi v^2}\mathbf{V}(x)\;\mathbf{Y}
   \left( \frac{1}{1+\frac{i \omega}{2 \pi v^2}\mathbf{C}_{s}\mathbf{Y}}\right) \tilde{\mathbf{V}}(y) \,,
\end{equation}
with $\mathbf{V}(x)=\int dz  \;\mathbf{C}(x|z)e^{\frac{i\omega z}{v}} $ and $\tilde{\mathbf{V}}(y)=\int dz  \;\mathbf{C}(z|y)e^{\frac{i\omega z}{v}} $. Instead of the full form \eqref{L_solution}, we can use  its simplified form integrated over the coordinates similarly to Eq.~\eqref{C_simple}
\begin{equation}
    \mathbf{L}^R_{s}=\left( \frac{1}{1+\frac{i \omega}{2 \pi v^2}\mathbf{C}_{s}\mathbf{Y}}\right) \mathbf{C}_{s}
    \,.
\end{equation}
The quantity $L^{R}_{\omega}(m|k)$ introduced in \eqref{eq:L_simple_1}  and entering  the equation for currents \eqref{J_full_equation}, is simply given by the corresponding matrix element $(\mathbf{L}^R_{s})_{mk}$. Schematically, this formula has the same structure as reported in \cite{Aristov2014} for the case $\alpha=0$, although the main difference lies in the concrete form of $\mathbf{C}_{s}$.

The quantity $L^{R}_{\omega}(m|k)$  allows the decomposition
  \begin{equation}
    \mathbf{L}_{s}^R
   =-\frac{2 i \pi v^2}{\omega}\mathbf{U}_3\left( \frac{1}{\mathbf{P}^{-1}+\mathbf{U}_3\mathbf{Y}\mathbf{U}_3}\right) \mathbf{U}_3\,, 
\label{LsR}
\end{equation}
where we diagonalized $\mathbf{C}_s$ by means of the unitary transformation
 \begin{equation}
\begin{aligned}
      \mathbf{U}_3 & =\frac{1}{\sqrt{2}}\begin{pmatrix}
    1 & 1 & 0\\
    1 & -1 & 0\\
    0 & 0 & \sqrt{2}
    \end{pmatrix},  \\
 \mathbf{P}&   = \frac{i\omega}{2\pi v^2}  \mathbf{U}_3 \mathbf{C}_s \mathbf{U}_3 
 = \mbox{diag }[P_{1},P_{2},P_{3}]\,.
\end{aligned}
\end{equation}
From Eqs.~ \eqref{C_solution} and  \eqref{J_v_trick} we obtain the explicit form of $P_{i}$:
\aleq{\label{C_eigen}
P_i=\frac{1}{2v^2\tilde{d}^2}\sum\limits_{\sigma, g, s = \pm} \kappa_\sigma \tilde{\theta}_2^{vv_s}\tilde{\chi}_1^{v,-gv_\sigma}\Big(\sum\limits_{j}b_{sj}^{\sigma g} -\delta_{s\sigma}\delta_{g+}\Big)\,,
}
where $\tilde{\theta}_2^{vv_s}$ is defined in Appendix~\ref{b_solution}.
The only $\omega$-dependence of $P_{i}$  comes from coefficients $b_{sj}^{\sigma g}$ containing  oscillatory exponents  $e^{\pm i\omega L/v_{\pm}}$ both in their numerator and denominator.

\section{RG equations for conductances}\label{sec:RG_eq_strong}
Let us briefly summarize the progress that we have done so far. The main result of the section \ref{ladder_section} is the resummation of an infinite series of relevant scale-dependent contributions to the bare bosonic propagator $L^{R,(0)}_{\omega}$ introduced
in \eqref{L0_bare}. The most general expression for this quantity is given in \eqref{L_solution}.

\begin{figure}
\includegraphics[width=0.9\columnwidth]{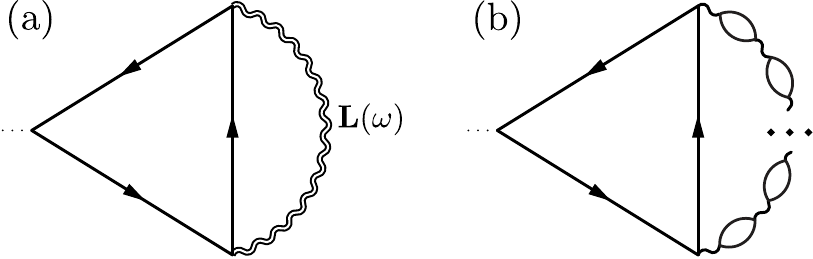}
\caption{(a) The diagram leading to the current correction due to the self-energy renormalized interaction. Double-wavy line is the solution of the integral equation \eqref{WH_1} depicted in Fig.~\ref{fig:eq_L}. (b) The typical diagram included in \eqref{current_full} in terms of bare interaction propagators (wavy lines) and fermionic loops.
}
\label{Fig:Full_triag}
\end{figure}

Now let us make use of the general formalism described in Sec.~\ref{sec:first_order} in order to extend the perturbative treatment of corrections to the currents into the strong coupling regime. Specifically, we substitute the bare propagator $L^{R,(0)}_{\omega}$ in \eqref{J_full_equation} by its fully-dressed version which corresponds to the self-energy renormalized
one-loop contribution. The resulting expression takes the form 
\aleq{\label{current_full}
    J_{j}= \sum_{mkp} \Re[ M^j_{mkp}]  \int\limits_{\epsilon}^{\omega_0} \frac{d\omega}{\omega}  F(\omega,V_{mp}) \\
     \times \Re\left( \mathbf{U}_3\left( \frac{1}{\mathbf{P}^{-1}+\mathbf{U}_3\mathbf{Y}\mathbf{U}_3}\right) \mathbf{U}_3\right)_{mk}\,,
}
where we took into account that $\mathbf{P}$ depends on frequency $\omega$ only through exponents and, thus, $\mathbf{P}(-\omega)=\mathbf{P}^{*}(\omega)$. All other parts of this equation were defined in Sec.\ref{sec:first_order}. Diagrammatically, this correction can be represented as Fig.~\ref{Fig:Full_triag}(a). Accordingly, the typical diagram included in this resummation is depicted in Fig.~\ref{Fig:Full_triag}(b).

We want to evaluate the integral in \eqref{current_full} with logarithmic accuracy in the limit $L\rightarrow \infty$ and at $T=0$. The most problematic part here is related to the incommensurate oscillations in $\mathbf{P}$ with several characteristic frequencies. It implies that the averaging over one period of oscillations, presented in \cite{Aristov2014} for a single characteristic frequency $v/L$, is not useful in our case since $\mathbf{P}$ is not truly periodic.

However, this difficulty can be overcome by shifting the contour of integration in the upper half-plane $\omega \rightarrow \omega + i \delta$, with $\delta \agt v/L \to 0$. Upon this deformation we do not encounter any poles in the upper half-plane of complex $\omega$ in view of the retarded nature of propagator and of the possibility to use the analytic digamma function $\psi[-ix/2\pi T]$ instead of $\coth(x/2T)$ in $F(\omega, V)$ in Eq.~\eqref{eq:L_simple_1}. 

More precisely, we note that Eq.\ \eqref{current_full} initially contained the integration over negative and positive $\omega$. The odd-in-$\omega$ property of $F(\omega, V)$ leads to picking the odd-in-$\omega$ component of $\mathbf{L}_{s}^R$,  Eq.\eqref{LsR}. The oddness of latter quantity allows one to add an even function to $F(\omega, V)$ without changing the result of integration. 
On the real axis of  $\omega$ we use  the identity $x \coth (x/2T) = 1/2+ x \Im \psi [-ix/2\pi T]$ and eventually replace $\Im \psi[-ix]$ by  $\psi[-ix]$, because $\Re\psi[-ix]$ leads to even-in-$\omega$ term in $F (\omega, V)$ which is integrated to zero.   

By examining \eqref{C_eigen} we notice that upon this shift, $\omega \rightarrow \omega + i \delta$,  the dominant contribution to $\mathbf{P}$ in the denominator of \eqref{current_full} will be determined by terms with the exponents $e^{-i \omega L (1/v_- +1/v_+)}$, acquiring additional factor $e^{ L \delta (1/v_- +1/v_+)} \gg 1$. The overall factor $ \omega^{- 1} $ gives a logarithmic divergence at low energies, which is regulated by $F(\omega,V_{mp})$ with the voltage $|V_{mp}|$ acting as an infrared cutoff scale. 

Therefore, the leading logarithmic divergence is simply given by \eqref{current_full} with the following replacement

\aleq{\label{P_eigen_reduced}
P_i\rightarrow\tilde{P}_i= \frac{1}{2v^2\tilde{d}^2}\sum\limits_{\sigma s} \kappa_\sigma \tilde{\theta}_2^{vv_s}\tilde{\chi}_1^{v,-v_\sigma}\Big(\tilde{b}^{\sigma +}_{s+}-\delta_{s\sigma} \Big)\,,
}
and the only reduced matrix elements which survived the procedure described above are given as $\tilde{b}^{+ +}_{++}=(\overline{\tilde{F}}_{1})^{-1}\overline{F}_{1}$, $\tilde{b}^{+ +}_{-+}=(\overline{\tilde{F}}_{1})^{-1}F_3$, $\tilde{b}^{- +}_{++}=-(\overline{\tilde{F}}_{1})^{-1}\left.F_{3}\right|_{v_{+} \rightarrow v_{-}}$, and $\tilde{b}^{-+}_{-+}=(\overline{\tilde{F}}_{1})^{-1}\tilde{F}_1$ (see App.~\ref{b_solution} for definition of $F_i$). 
After some algebra one can obtain the following transparent form of these eigenvalues
\begin{equation}\label{P_as_K}
    \tilde{P}_1 =  \frac{\left.\mathcal{K}\right|_{\tau=0}-1}{\left.\mathcal{K}\right|_{\tau=0}+1}\,,\quad\tilde{P}_2 =  \frac{\mathcal{K}-1}{\mathcal{K}+1}\,,\quad \tilde{P}_3 = \frac{K_3-1}{K_3+1}\,,
\end{equation}
where $K_3=(1+2g_3)^{-1/2}$, and we introduced the modified 
Luttinger parameter
\begin{equation}\label{eq:kcal}
   \mathcal{K}=K\left[\frac{\tau(\tilde{K}+\xi )+(1-\tau)W \tilde{K}}{\tau W K+(1-\tau)(K+\xi \tilde{K}/K)}\right]\,,
\end{equation}
which is expressed in terms of original Luttinger parameters
\begin{equation}
    K=\frac{1}{\sqrt{1+2g}},\quad \tilde{K}=\frac{1}{\sqrt{1+2g-\alpha}}\;,
\end{equation}
where $\tilde K$ is the Luttinger parameter in the absence of retardation effects, which formally corresponds to the limit of the infinite phonon velocity with $\xi=0$. We also defined one extra combination 
$W=(v_++v_-)/c$ which can be represented as 
\begin{equation}
    W=\sqrt{\left(1+K^{-1}\xi\right)^2+2\xi\left( \tilde{K}^{-1}-K^{-1} \right)}.
\end{equation}
It will be convenient to use a different parametrization of the form 
\aleq{\label{defq}
\tilde{\mathbf{P}}=-\diag\left\{q_1^{-1},q_2^{-1},q_3^{-1} \right\}
}
in order to match the notation introduced in \cite{Aristov2009}.

Thus, we only need to evaluate the logarithmic divergence in the remaining integral over frequency identical to that we discussed in \eqref{log_integral}.

Finally, we obtain
\aleq{\label{Current_full_log}
       J_j  = 2\sum\limits_{mkp} \left(\mathbf{U}_3\left(\frac{1}{\mathbf{\tilde{P}}^{-1}+\mathbf{U}_3 \mathbf{Y}\mathbf{U}_3} \right)\mathbf{U}_3 \right)_{mk}
       \\\times V_{mp} \operatorname{Re}[M_{mkp}^j]\; \ln \left(\frac{\omega_0}{\rm{max}\{|V_{mp}|,\epsilon\}}\right)\,,
}
where $\mathbf{\tilde{P}}$ is given in \eqref{P_as_K}.

We note that in case of finite temperatures the large logarithms remain finite even for $\epsilon=0$, as
the low-energy cutoff is provided by the function $F(\omega,V)$
taken at finite $T$. Then \eqref{log_integral} should be replaced by $\mathcal{I}\left(\omega_{0}, T, V\right) \approx 2 V \theta\left(T-c_{*}|V|\right) \ln \frac{\omega_{0}}{2 \pi T}$, where $c_{*}$ is a number of the order of unity given in \cite{Aristov2017}.

The corrections to the currents, Eq.~\eqref{Current_full_log},  translate into the corresponding corrections for conductances~\eqref{currents_conductances}. The conductances scaling hypothesis allows us to write non-perturbative RG equations in the same way as in the first order case (see Section \ref{sec:first_order}):
\aleq{
\label{RGeq3g}
   \frac{d G_a}{d\Lambda} &= 2A_1 \theta_a(\epsilon)+A_2 \theta_+(\epsilon),\\
    \frac{d G_b}{d\Lambda} &= 2B_2 \theta_+(\epsilon),
}
where the strong-coupling counterparts of \eqref{A,B_weak} are given by
\aleq{\label{eq:AB}
A_1&=-2\frac{G_a(1-G_a)-G_b/4}{q_2-(1-2G_a)},\\
A_2&=-\frac{G_b}{8}\left[\frac{1}{q_2-(1-2G_a)}+\frac{4(1-2G_a)}{Q-(1-2G_b)} \right],\\
B_2&=-\frac{G_b}{4}\left[\frac{1-2G_a}{q_2-(1-2G_a)}+\frac{4(1-G_b)}{Q-(1-2G_b)} \right],
}
and we defined (compare  with  notation in \cite{Aristov2011})
\aleq{\label{eq_Q}
Q=\frac{4 q_1 q_3-2 q_1-3 q_3+1}{2 q_1+q_3-3}.
}
We note that in the fully non-equilibrium setting the electron-phonon coupling constant $\alpha$ encoded in \eqref{eq:AB} is assumed to be proportional to the step function  $\theta_D(\epsilon)=\theta(\omega_D-\epsilon)$, in agreement with our convention introduced in Sec.~\ref{sec:III_first_order} B.

General beta functions \eqref{RGeq3g},\eqref{eq:AB} derived for arbitrary parameters of our model (such as electron-phonon coupling constant, coefficients of $\mathbf{B}$ and $\mathbf{S}$ matrices, etc.) describe renormalization of dc conductances due to interaction effects. This is the central result of our derivation. In the next section we will discuss physical implementations of these RG equations in different limiting cases.

It is worth to comment here on the completeness of the derived set of RG equations \eqref{RGeq3g}. Based on the intuitive similarity of the fermionic $S$ matrix and the phonon $B$ matrix, one might ask if it is necessary to construct analogous RG equations for the $B$ matrix as well. Somewhat related questions were raised in \cite{Yurkevich2013}, where it was proposed that phenomenologically imposed correlations between fermion and phonon scattering matrix elements could potentially lead to the existence of new unstable RG fixed points. In our formalism, we didn't find any contributions to the phonon propagator that could be interpreted as separate scale-dependent corrections of the phonon $B$-matrix. Logarithmic corrections to the phonon propagator emerge from diagrams containing fermion loops, they correspond to iteration of RG equation for the $S$ matrix together with the above RPA-type summation. Hence, no additional scaling equations for the $B$-matrix are required.

\section{Scaling exponents at strong coupling}\label{sec:scaling_exp_strong}

The above calculation provided us with the non-equilibrium strong-coupling RG equations for conductances. As we show now, they are in agreement with all previously known limiting cases: electron-electron strong coupling in equilibrium regime for Y-junction~\cite{Aristov2011},  electron-electron weak coupling in non-equilibrium regime for Y-junction~\cite{Aristov2017},  electron-electron strong coupling in non-equilibrium regime for impurity case (when the tip is absent)~\cite{Aristov2014}, electron-phonon strong coupling in equilibrium regime for impurity case near the conductances fixed points~\cite{Yurkevich2013}.

Non-equilibrium RG equations in the first order of interaction were discussed to some length in Sec.~\ref{sec:first_order}. The sophisticated summation of Sec.\ \ref{ladder_section} does not qualitatively change the above off-equilibrium picture. So let us to focus instead on the equilibrium case: $V_{a,b} \to 0$ and $\omega_D \to \infty$, where non-perturbative treatment of interactions reveals new phenomenon.

\subsection{Wire with an impurity}

\begin{figure}[h!]
 \center{\includegraphics[width=0.9\linewidth]{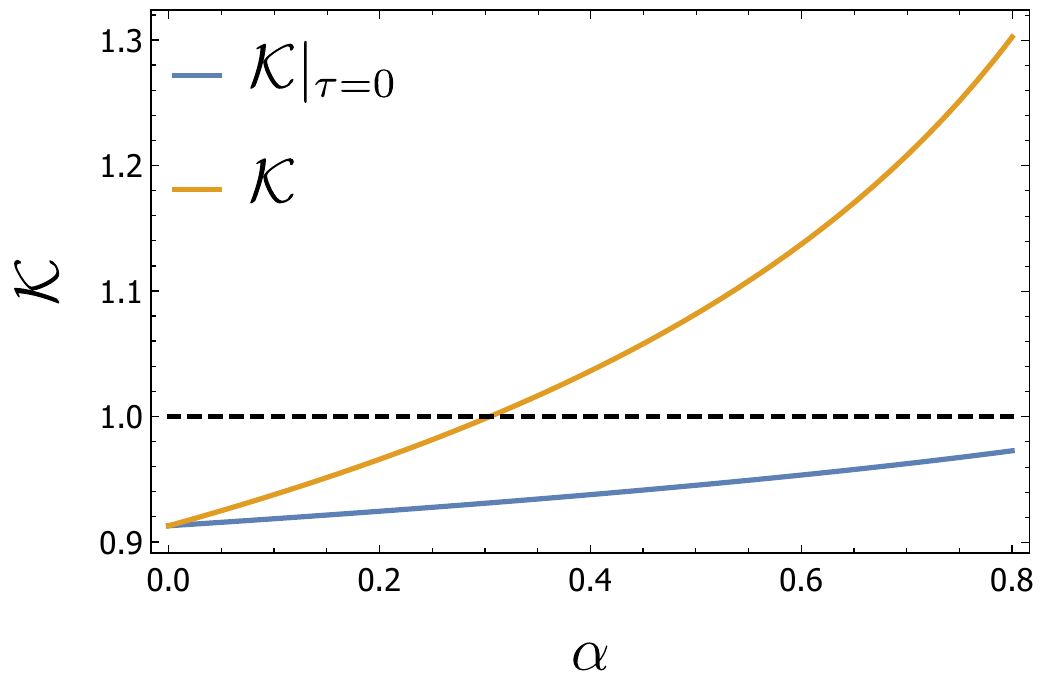}}
\caption{The modified Luttinger parameter $\mathcal{K}$ for $\tau=0$ (blue line) and for $\tau=1$ (orange line) for $g=0.1$ and $\xi=1.5$.}
\label{fig:mathcal_K}
\end{figure}

Earlier, the case of the wire with the impurity was studied by the bosonization technique in presence of electron-phonon interaction~\cite{Yurkevich2013}. For Y junction this means the decoupled tip: $t_2=0$, $r_2=1$ for S matrix~\eqref{smatrix}, i.e. $G_b=0$~\eqref{eq:barecond}.
The only RG equation in this case is 
\begin{equation}
    \frac{dG_a}{d\Lambda}=2A_1=-\frac{4G_a(1-G_a)}{q_2 - (1-2G_a)}.
\end{equation}
Only two fixed points exist: N point, $G_a=0$, and A point, $G_a=1$. In the vicinity of these points the linearized equations have the standard form    
\aleq{\label{RG_wire_strong}
\frac{dG_a}{d\Lambda} &=2(1-\mathcal{K}^{-1})G_a\,, \quad& G_a&\ll  1\,,\\
  \frac{d\tilde{G_a}}{d\Lambda}&=2(1-\mathcal{K})\tilde{G_a}\,, \quad &\tilde{G_a}&=1-G_a\ll 1 \,.
}
This result is in exact correspondence with Yurkevich et al.\ ~\cite{Yurkevich2013}.
The fixed point N corresponds to the total loss of conductance. It is stable for $\mathcal K < 1$, while the fixed point A corresponds to the ideal transmission case and is stable in the opposite situation, $\mathcal K > 1$. The renormalization is absent for $\mathcal K=1$, though the coupling constants might not be zero in this case, see Fig.~\ref{fig:mathcal_K}.
Another interesting point is that in a certain range of parameters the modified Luttinger parameter can be $\mathcal{K} < 1 $ for $\tau=0$, while by increasing $\tau$ one can continuously increase $\mathcal{K}$ up to the values greater than one.

Without the electron-phonon interaction, $\alpha=0$, one has the well-known result $\tilde{K}=K$ with the scaling exponents $K$ and $1/K$ for the weak scattering and weak link limits respectively. In this case $K<1$ for the repulsive interaction between electrons, $g>0$, and $K>1$ for the attractive electron-electron interaction, $g<0$.

Let us also discuss the role of the kinetic asymmetry between bosonic and fermionic modes. For $\xi=0$ (phonon velocity tending to infinity) we get $\mathcal{K}=\tilde{K}$ which corresponds to the absence of any retardation effects (purely screened local interactions). In the opposite case $\xi\rightarrow\infty$ (phonon velocity tending to zero) one obtains $\mathcal{K}=K$ and phonons are incapable to modify renormalization. In the intermediate regime $\mathcal{K}$ is a monotonic function of $\xi$.

\subsection{Y-junction}

In the equilibrium limit the strong coupling RG equations read as
\aleq{
\label{RGeq2g}
   \frac{d G_a}{d\Lambda} &= 2A_1+2 A_2 ,\\
    \frac{d G_b}{d\Lambda} &= 4B_2,
}
with the above definition~\eqref{eq:AB}.
Two universal fixed points of this RG system, $N$ and $A$, can be characterized by two independent scaling exponents corresponding to different directions in the space of conductances, $(G_a,G_b)$: along the line $G_b=0$, and along the boundary  $G_b/4=G_a(1-G_a)$. In addition to above fixed  points at the line $G_b=0$ (corresponding to the detached tip), the saddle-type fixed point $M$ appears. As discussed in Sec. \ref{sec:first_order} the position of the latter point $M$ at the parabola of allowed conductances is not universal. In this case, the associated scaling exponents are naturally related to two directions, one along the parabola and another perpendicular to it. Their exact form is rather cumbersome and given by Eq.(33) and (35) in \cite{Aristov2011} where parameters $q$ and $Q$ (which is a function of $q$ and $q_3$) should be replaced by our expressions for $q_2$ and $Q$ (as a function of $q_1$ and $q_3$), given by \eqref{defq} and \eqref{eq_Q} respectively.

 The first set of exponents for the fixed points $N$ and $A$ can be read from \eqref{RG_wire_strong} and is given by
\aleq{
\gamma_{N,1}&=2(\mathcal{K}^{-1}-1)\,,\\
\gamma_{A,1}&=2(\mathcal{K}-1)\,.
}
The second set of scaling exponents is found as
 \begin{equation}
\begin{aligned}
    \gamma_{N,2} &=K_3^{-1}+\frac{1}{2}(\left.\mathcal{K}^{-1}\right|_{\tau=0}+\mathcal{K}^{-1})-2 ,\\
    \gamma_{A,2} &=K_3^{-1}+\frac{1}{2}( \left.\mathcal{K}^{-1}\right|_{\tau=0}+\mathcal{K})-2 
\end{aligned}
\label{eq:ScalExp}
\end{equation}
and additional exponents defined in \eqref{G_c_eq} and \eqref{G_b_eq} can be obtained as 
$\kappa_{p,j}=\left.\gamma_{p,j}\right|_{\alpha=0}-\gamma_{p,j}$ with $p=N,A$ and $j=1,2$. In the weak coupling regime, these scaling exponents coincide with the ones presented in Sec.~\ref{sec:fp1}.

\begin{figure}[h!]
 \center{\includegraphics[width=0.98\linewidth]{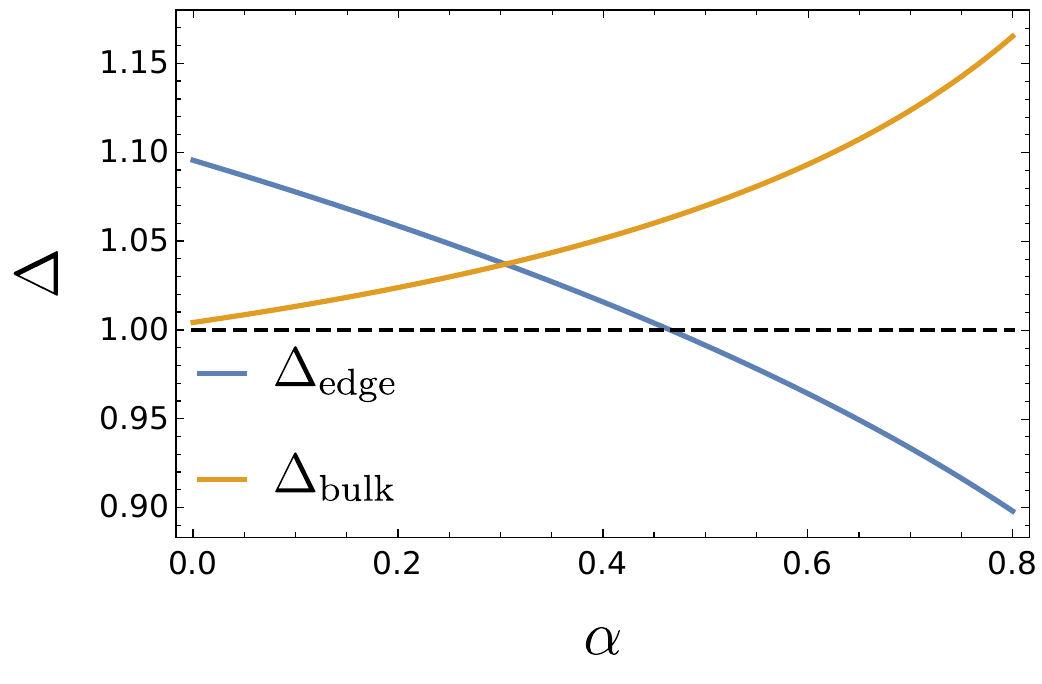}}
\caption{The boundary exponent $\Delta_{\text{edge}}=\frac{1}{2}(\left.\mathcal{K}^{-1}\right|_{\tau=0}+\mathcal{K}^{-1})$ (blue line), and the bulk exponent  $\Delta_{\text{bulk}}=\frac{1}{2}( \left.\mathcal{K}^{-1}\right|_{\tau=0}+\mathcal{K})$ (orange line) for $g=0.1$, $\tau=1$ and $\xi=1.5$. The fixed point $M$ exists when $\Delta_{\text{edge}}>1$ and $\Delta_{\text{bulk}}>1$. }
\label{fig:Delta_exponents}
\end{figure}

Let us discuss the expressions \eqref{eq:ScalExp}. The fixed point $N$ is characterized by the exponent $\gamma_{N,2}$,   determined by the sum of the boundary exponent for the third wire $K_3^{-1}$ and the new combination $\Delta_{\text{edge}}=\frac{1}{2}(\left.\mathcal{K}^{-1}\right|_{\tau=0}+\mathcal{K}^{-1})$. We will refer to this combination as an effective boundary exponent for the main wire. The fixed point $A$ has the exponent $\gamma_{A,2}$, which is controlled by the boundary exponent of the third wire $K_3^{-1}$ and the effective bulk anomalous dimension of the fermion operator $\Delta_{\text{bulk}}=\frac{1}{2}( \left.\mathcal{K}^{-1}\right|_{\tau=0}+\mathcal{K})$. This quantity corresponds to the well-known zero-bias anomaly and controls the suppression of the tunneling density of states \cite{Aristov2010}. Both exponents are depicted on Fig.~\ref{fig:Delta_exponents} as a function of the electron-phonon coupling constant $\alpha$. 
We note in passing that the condition for the existence of $M$ point reads as $\gamma_{N,2}\gamma_{A,2}>0$. 

The way how two modified Luttinger parameters $\left.\mathcal{K}^{-1}\right|_{\tau=0}$ and $\mathcal{K}$ enter scaling exponents can be understood as follows. The first term in expressions for $\Delta_{\text{edge}}$ and $\Delta_{\text{bulk}}$ is always associated with the direct tunneling processes from one of the arms of the main wire to the tip, and thus, has a characteristic form of the weak link exponent, i.e. is inversely proportional to the Luttinger parameter. Additionally, since in our model we assume the absence of phonon transport between the main wire and the tunneling tip, then undergoing this process fermions are not affected by the interwire interactions associated with the transmission coefficient $\tau$, which in turn results in $\left.\mathcal{K}^{-1}\right|_{\tau=0}$ for both fixed points. The second contribution to the boundary and bulk exponents corresponds to scattering processes within the main wire and has a tunneling or weak-scattering form for $N$ and $A$ points respectively. However, in both cases propagating fermions can interact through the exchange of phonons across a junction, and consequently, this term comes fully dressed with non-zero $\tau$.

An important statement is the following. Contrary to what was obtained in the previous studies of Y junctions  in the absence of  phonons, the scaling exponents now can not be written in terms of a \emph{single} modified Luttinger parameter. This fact may have consequences for possible attempts to recover the strength of interaction from two experimentally observed exponents, $\Delta_{\text{edge}}$ and $\Delta_{\text{bulk}}$.  
In the anticipated situation of non-interacting tip, $K_{3}=1$,   one can naively extract the effective Luttinger parameter from two alternative  definitions \cite{nazarov_blanter}
\begin{equation}\label{eq:Keff_def}
    K_{\text{eff}}^{\text{(edge)}}=\frac{1}{\Delta_{\text{edge}}}\,, \quad \frac{1}{2}\left(\frac{1}{K_{\text{eff}}^{\text{(bulk)}}}+K_{\text{eff}}^{\text{(bulk)}} \right) =\Delta_{\text{bulk}}\,.
\end{equation}

\begin{figure}
 \center{\includegraphics[width=0.98\linewidth]{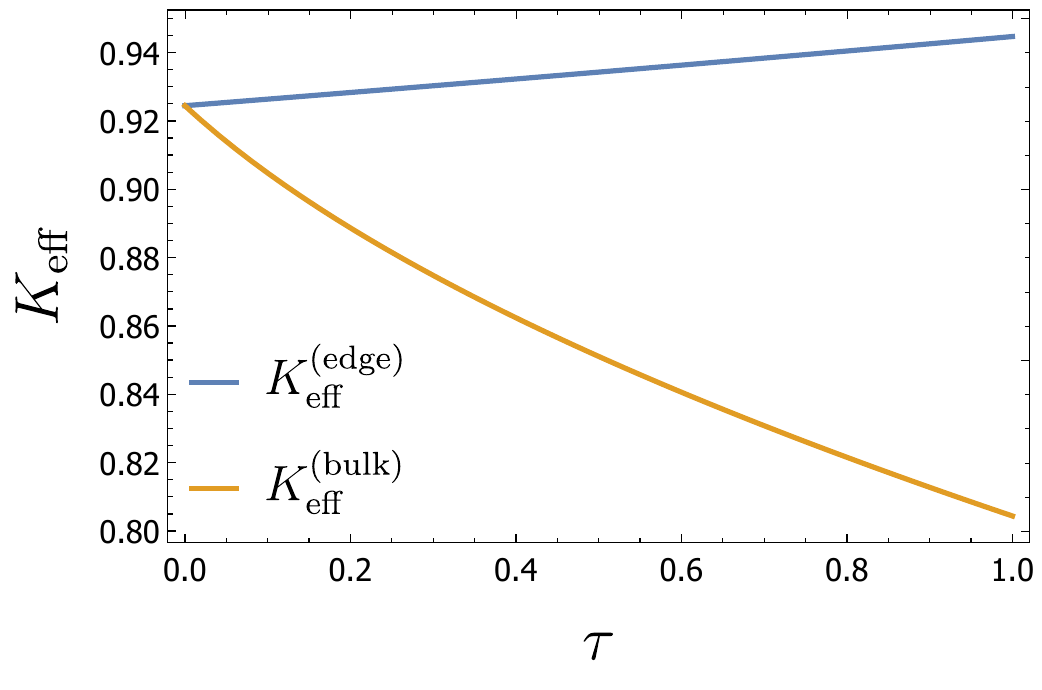}}
\caption{Two effective Luttinger parameters $K_{\text{eff}}^{\text{(edge)}}$ (blue line)  and $K_{\text{eff}}^{\text{(bulk)}}$ (orange line) (see Eq.~\eqref{eq:Keff_def}) as a function of $\tau$ for $\alpha=0.2$, $g=0.1$ and $\xi=1.5$. They don't coincide for $\tau>0$, contrary to naive expectation following from all previously known Y-junctions studies.}
\label{fig:Keff}
\end{figure}

These effective Luttinger parameters are depicted on Fig.~\ref{fig:Keff} as a function of $\tau$. They coincide only for the phonon ideal reflection case, $\tau=0$. On the other hand, in the tunneling experiments where the phonons pass through the vicinity of Y junction, $\tau \neq 1$, our formulas \eqref{eq:ScalExp} show that in the physically relevant range of parameters $K_{\text{eff}}^{\text{(bulk)}}$ is smaller than $K_{\text{eff}}^{\text{(edge)}}$, and thus, 
  $K_{\text{eff}}$ should be determined differently.

In fact, in certain experimentally studied  low-dimensional systems exhibiting LL-type behaviour the similar mismatch between measured Luttinger parameters  was observed. For instance, in artificial atom chains \cite{blumenstein2011} $ K_{\text{eff}}^{\text{(edge)}}$ extracted from experimental data was found to be larger than $ K_{\text{eff}}^{\text{(bulk)}}$, in agreement with our predictions. 
Our calculation shows that the difference between two Luttinger parameters \eqref{eq:Keff_def} can be at least partially due to the strong electron-phonon interaction.

Instead of a single effective Luttinger parameter we propose to characterize tunneling experiments in Luttinger Liquids by using both $\left.\mathcal{K}^{-1}\right|_{\tau=0}$ and $\mathcal{K}$ which can be determined from observable boundary and bulk exponents as follows
\aleq{
\left.\mathcal{K}^{-1}\right|_{\tau=0}&=\Delta_{\text{bulk}}+\Delta_{\text{edge}}-\sqrt{1+(\Delta_{\text{bulk}}-\Delta_{\text{edge}})^2}\;,\\
\mathcal{K}&=\Delta_{\text{bulk}}-\Delta_{\text{edge}}+\sqrt{1+(\Delta_{\text{bulk}}-\Delta_{\text{edge}})^2}\;.
}

Thus, experimental measurements of the conductances scaling behaviour near the two fixed points $N$ and $A$ (i.e. for two different junction regime: ideal reflection and transmission case, correspondingly) allow to obtain 
interwire interactions contribution to the electron transport in the junction.

It worth noting that the boundary and bulk exponents can be equal to each other on some non-trivial surface in the parameter space (for $\tau=0$ or without phonons it can happen only in the non-interacting case $K=1$). This situation is accompanied by the $M$ point being located exactly on the top of the parabola of allowed conductances and the emergence of the line of fixed points located at $G_b=0$ as was discussed in Sec.~\ref{sec:first_order}. We note that these two exponents equal to each other ($\Delta_{\text{edge}}=\Delta_{\text{bulk}}$) exactly at the point where the condition  $\mathcal{K}
=1$ is satisfied.

It is straightforward to show that scaling exponents presented in this section can be easily generalized to account for the additional electron-phonon interaction present in the third wire ($\alpha_3\neq 0$) with the same $B$-matrix \eqref{eq:Bmat}. Physically, it corresponds to the situation when an ideal tunneling tip is replaced by an electrode made from the same material as a main wire with large electron-phonon coupling. The Luttinger parameter $K_3$ then should be simply replaced by $\mathcal{K}_3$ with $\tau=0$ (since the corresponding matrix element is $B_{33}=1$).

Finally, we note that although the scaling exponents derived within our approach are significantly modified in the presence of the electron-phonon interaction, the RG equations \eqref{RGeq2g} {\it do not} exhibit any new fixed points in addition to already described cases \cite{Aristov2011}, see also \cite{Yurkevich2013}. It can be understood in terms of the full integral equation \eqref{WH_0}. The bare interaction propagator plays a role of the ``starting point" for the RPA dressing procedure (see Sec.~\ref{ladder_section}). If one starts with just a local interaction, then already the first iteration of the integral equation results in the new kernel with the form structurally resembling the phonon propagator where $B$ matrix is replaced by matrix elements of the fermionic $S$ matrix. Thus, away from $N$ and $A$ fixed points, then conductances are finite, this decoration of the interaction potential effectively ``smoothens" the difference between initial bare bosonic propagators. As a result, the standard classification of fixed points applies. On the other hand, in the proximity of the fixed point $N$ the corresponding tunneling matrix element $t_2$ renormalizes to zero, and the terms proportional to $\tau$ in \eqref{L_solution} are the only non-diagonal contributions that survive and drastically change scaling exponents.

\section{Conclusions}
\label{sec:diss}
In this work, we have studied the effect of the electron-phonon interaction on the renormalization of conductances in the Y-junction of the Luttinger liquids out of equilibrium. This problem setup corresponds to the geometry of a scanning tunneling microscopy experiment of one-dimensional quantum systems, for example, carbon nanotubes \cite{Izumida2005} or helical edge states of topological insulators \cite{Das2011,Stuhler2019}.

Within the fermionic approach enforced by the Keldysh diagrammatic technique, corrections to charge currents were calculated at the infinite order of the perturbation theory in the electron-phonon coupling constant, and scale-dependent logarithmic contributions were determined. This allowed us to apply the renormalization group formalism and derive the beta functions at strong coupling for two characteristic conductances $ G_a $ and $ G_b $, which correspond to the current in the main wire and the tunneling tip, respectively. The obtained renormalization-group equations were solved analytically in the vicinity of the fixed RG points, and the corresponding scaling exponents, as well as various non-equilibrium regimes, were analyzed in details.

 The system exhibits two typical transport behaviors in correspondence with two possible  fixed points for the RG flows in the plane of conductances. When the attractive electron-phonon interaction is small enough in comparison with the repulsive Coulomb interaction then the conductance $G_a$ tends to zero (fixed point $N$, ``insulator''  behavior), in opposite case for certain parameters $G_a$ tends to ideal conductance value (fixed point $A$, ``metal''  behavior). Additionally, there is a saddle-type fixed point M.
The physical reason for the appearance of the non-universal $ M $ point in the absence of electron-phonon interaction was the competition between the renormalization of the tunneling density of states and the instability with respect to the formation of a charge density wave in the main wire \cite{Aristov2010}. 
We show that the influence of phonons on $M$ point is two-fold. One effect is induced attraction between electrons, which in the absence of interaction in the tip would lead to disappearance of $M$ point, see Fig.\ 4 in \cite{Aristov2011}.
Another effect  favors the existence of $M$ point and concerns the non-locality of phonon-mediated interaction, in particular the interaction over the barrier, $\tau>0$. We emphasize that, although the repulsive interaction in the tunneling tip, $ g_3 $, leads to the appearance of $ M $ point \cite{Aristov2017}, it cannot move the fixed point $ M $ to the left half of the RG diagram. In  case of only two quantum wires with an impurity (limit $ G_b = 0 $), the calculated scaling exponents for fixed points coincide with those reported in \cite{Galda2011} within the bosonization framework.

We also demonstrated that the presence of an additional ultraviolet scale in the model, determined by the Debye frequency $ \omega_D $, enriches non-equilibrium transport regimes. As a result, rather complicated RG trajectories may  exist: conductances RG flow  can change the direction from the fixed point $N$ (insulating  behavior) to the fixed point $A$ (metallic  behavior) with running energy $\epsilon$ decreasing due to the non-universal position of the $ M $ point. Specifically, at high energies (temperature or voltages greater than the Debye frequency $ \omega_D $), the contribution originating from the electron-phonon coupling is irrelevant, and the $ M $ point is located on the right side of the RG diagram. At energies lower than the Debye scale, the contributions of inelastic scattering with phonon transfer begin to play a crucial role in renormalization. If the junction is transparent for phonon transport $ \tau > 0 $, then the tunneling density of states is suppressed, and the $ M $ point changes its position and affects the directions of RG flows (see Fig.~\ref{fig:NonEqDebye}). As a result, the dependence of conductances on the infrared cutoff (for example, temperature) turns out to be non-monotonic.

Finally, we show that the scaling of conductances of Y junction is governed by two effective Luttinger parameters, related to the main wire. For the geometry of the Luttinger liquid wire with impurity (detached tunneling tip) only one Luttinger parameter, $\mathcal{K}$, appears in equations. Rather unexpectedly, the scaling exponents for the tunneling tip conductance are defined by both the previous $\mathcal{K}$, and $\mathcal{K}|_{\tau=0}$, calculated in geometry of Y junction impenetrable for phonons. It means that the Luttinger parameter, $\mathcal{K}$, naively extracted from the bulk tunneling exponent of tip conductance will show systematic deviation, due to phonons, from $\mathcal{K}$, determined in other types of experiment.

\begin{acknowledgments}
The work of R.N. and P.N. was funded by RFBR according to the research project No. 18-32-00424. The work of D.A. was funded by RFBR and DFG according to the research project No. 20-52-12019. Also, the work of R.N. was partly supported by the grant of the Foundation for the Advancement of Theoretical Physics ``BASIS''.
\end{acknowledgments}

\appendix

\section{Phonon propagator with a single impurity}
\label{AppendixA}
For simplicity we consider a mass defect model of impurity, while similar calculations can be performed for the pinning or elastic defects~\cite{San-Jose2005}.
We introduce the causal Green's function for the lattice deformations $u(x)$. In the $(\omega,x)$ representation it reads
\begin{equation}
U^0(\omega,x,x')=-\frac{i}{2c|\omega|}e^{i\frac{|\omega|}{c}|x-x'|}\,.
\end{equation}
We solve the Lippmann-Schwinger equation accounting for multiple phonon scattering.  The solution has the following form 
\begin{multline}
   U(\omega,x,x')= U^0(\omega,x,x')\\
   -\frac{\Delta m/m}{F(\omega)}\omega^2 U^0(\omega,x,0)U^0(\omega,0,x'),
\end{multline}
for $x,x'\neq 0$. The appearing pole corresponds to the localized vibrational mode
\begin{equation}
    F(\omega)=1+\frac{\Delta m}{m}\omega^2 U^0(\omega,0,0)=1-i\frac{|\omega|}{\omega_m}\,,
\end{equation}
with the characteristic frequency $\omega_m=2m c/\Delta m$.
In the main text we use the mixed wire and $(\omega,x)$ representation, so bare phonon propagator reads
\begin{equation}
    U^0_\omega(l,x|m,x')=-\frac{i}{2c|\omega|}\left(e^{i\frac{|\omega|}{c}|x-x'|}\sigma^0_{lm}+e^{i\frac{|\omega|}{c}(x+x')}\sigma^1_{lm} \right),
\end{equation}
for $x,x'>0$. The complete solution is 
\begin{multline}
    U_\omega(l,x|m,x')=-\frac{i}{2c|\omega|}\left(e^{i\frac{|\omega|}{c}|x-x'|}\sigma^0_{lm}\right. \\ \left.-\rho(|\omega|)e^{i\frac{|\omega|}{c}(x+x')}\sigma^0_{lm}+\tau(|\omega|)e^{i\frac{|\omega|}{c}(x+x')}\sigma^1_{lm} \right),
\end{multline}
and the reflection and transmission coefficients have the form 
\begin{equation}
    \rho(|\omega|) = -\frac{i|\omega|/\omega_m}{1-i|\omega|/\omega_m}\,, \quad \tau(|\omega|) = 1-\rho(|\omega|)\,.
\end{equation}
There are two simple limiting cases: first, if $\Delta m=0$ then we obtain $\rho =0$, $\tau=1$; and second, if $\Delta m = +\infty$ then we get $\rho =1$, $\tau=0$.

We notice the continuity condition at $x = 0$ of the Green function of deformations
\begin{equation}
U_\omega(1,x|1,0^+)=U_\omega(1,x|2,0^+)
\end{equation}
and for the full reflection case ($\rho=1$) we obtain the following boundary condition
\begin{equation}
U_\omega(1,x|1,0^+) =0 \quad \leftrightarrow \quad u(0)=0\,.
\end{equation}

For the electron-phonon interactions we need to consider the Green's function for the gradients of deformations
\begin{multline}
\tilde{D}(x,t,x',t')=-i\left< T c\nabla u (x,t) c\nabla u (x',t')  \right> \\
=c^2\partial_x\partial_{x'}U(x,t,x',t'), 
\end{multline}
for $x,x'\neq 0$. This definition can be reformulated in terms of our mixed wire and $(\omega,x)$ representation as
\begin{equation}
    \tilde{D}_{\omega}(l,x|m,x')=c^2(\sigma^0_{lm}-\sigma^1_{lm})\partial_x\partial_{x'}U_\omega(l,x|m,x'), 
\end{equation}
for $ x,x'>0$. Therefore, we obtain the following expression for the Green's function
\begin{multline}\label{GF_deformations}
      \tilde{D}_{\omega}(l,x|m,x')=-\delta(x-x')\sigma^0_{lm} \\-
     \frac{i|\omega|}{2c} \left(e^{i\frac{|\omega|}{c}|x-x'|}\sigma^0_{lm} +B_{lm}e^{i\frac{|\omega|}{c}(x+x')} \right)\,,
\end{multline}
here the phonon $\mathbf{B}$ matrix is an   analog of electron $|S_{ij}|^2$ matrix
\begin{equation}
    B=
\begin{pmatrix}
    \rho & \tau\\
    \tau & \rho
    \end{pmatrix}, \quad \tau+\rho=1\,.
\end{equation}
The retarded Green's functions is given by Eq.~\eqref{eq:greenphon} in the main text. 

\section{Useful operator identities for $D_{\rm{v}}$ }\label{App_operator_relations}
In this section we present several useful identities for differential operators $D_{\rm{v}}$ defined in the main text as $D_{{\rm{v}}}=\partial^2_x+\omega^2/{\rm{v}}^2$.

We are mostly interested in the result of action of $D_{\rm{v}}$ on various exponential functions because of the structure of our bare bosonic propagator in the mixed frequency-coordinate representation \eqref{L0_bare}. For instance, one can easily derive 
\aleq{\label{D_ident_1}
D_{\rm{v}}\; e^{i\frac{\omega}{V}x} &=\omega^2\left(\frac{1}{{\rm{v}}^2}-\frac{1}{V^2}\right)e^{i\frac{\omega}{V}x},\\
D_{\rm{v}}\; e^{i\frac{\omega}{V}|x|} &=\frac{2i\omega}{V}\delta(x)+\omega^2\left(\frac{1}{{\rm{v}}^2}-\frac{1}{V^2}\right)e^{i\frac{\omega}{V}|x|}\,.
}

Another important simplification is coming from the combination ${\cal D}_{2}=D_{v_+}D_{v_-}$ introduced in \eqref{diff_eq_C_operator}. By using eq.\eqref{D_ident_1} one can obtain
\begin{equation}\label{T_ident_1}
{\cal D}_{2}\; e^{i\frac{\omega}{V}x}=\omega^4\left(\frac{1}{v_-^2}-\frac{1}{V^2}\right)\left(\frac{1}{v_+^2}-\frac{1}{V^2}\right)e^{i\frac{\omega}{V}x}
\end{equation}
and   we get
\aleq{
{\cal D}_{2}\; e^{i\frac{\omega}{V}|x|}&=\frac{2i\omega}{V}\delta^{(2)}(x)+\frac{2i\omega^3}{V}\left(\frac{1}{v_+^2}+\frac{1}{v_-^2}-\frac{1}{V^2}\right)\delta(x)\\
&+\omega^4\left(\frac{1}{v_-^2}-\frac{1}{V^2}\right)\left(\frac{1}{v_+^2}-\frac{1}{V^2}\right)e^{i\frac{\omega}{V}|x|}
\,.}
It is useful to derive how ${\cal D}_{2}$-operator acts on a delta function:
\begin{equation}
{\cal D}_{2}\;\delta(x)=\delta^{(4)}(x)+\omega^2\left(\frac{1}{v_+^2}+\frac{1}{v_-^2}\right)\delta^{(2)}(x)+\frac{\omega^4}{v_-^2v_+^2}\delta(x).
\end{equation}
Now let us consider these relations for the specific case of $v_\pm$. From \eqref{T_ident_1} one can see that ${\cal D}_{2}$ vanishes on the propagating exponents $e^{i\frac{\omega}x{v_\pm}}$:
\begin{equation}
{\cal D}_{2}\; e^{i\frac{\omega}{v_+}x}={\cal D}_{2}\; e^{i\frac{\omega}{v_-}x}=0\,.
\end{equation}
Moreover, we obtain
\begin{equation}
{\cal D}_{2}\; e^{i\frac{\omega}{v_\pm}|x|}=\frac{2i\omega}{v_\pm}\delta^{(2)}(x)+\frac{2i\omega^3}{v_\pm v_\mp^2}\delta(x)\,.
\end{equation}
Finally, we consider the case with $V=c$:
\begin{equation}
{\cal D}_{2}\; e^{i\frac{\omega}{c}x}=\omega^4\left(\frac{1}{v_-^2}-\frac{1}{c^2}\right)\left(\frac{1}{v_+^2}-\frac{1}{c^2}\right)e^{i\frac{\omega}{c}x}\,,
\end{equation}
\aleq{
{\cal D}_{2}\; e^{i\frac{\omega}{c}|x|}=\frac{2i\omega}{c}\delta^{(2)}(x)+\frac{2i\omega^3}{c}\left(\frac{1}{v+^2}+\frac{1}{v_-^2}-\frac{1}{c^2}\right)\delta(x)\\
+\omega^4\left(\frac{1}{v_-^2}-\frac{1}{c^2}\right)\left(\frac{1}{v_+^2}-\frac{1}{c^2}\right)e^{i\frac{\omega}{c}|x|}\,.
}

\section{Boundary conditions for matrices $\mathbf{A}_{\sigma g}$}\label{reduction_A}

In this section we outline important steps in the derivation of the set of  equations for  matrices $A_{\sigma g}(y)$ \eqref{Full_Eq1}. As it was explained in the main text, we can use the ansatz \eqref{C_full} for the  integral equation \eqref{WH_equation_compact1} and compare coefficients corresponding to different linearly-independent $x$-functions. Thus, it is convenient to first evaluate integrals \eqref{integrals} in terms of matrices $\mathbf{A}_{\sigma g}$ by using \eqref{C_full}. We obtain
\aleq{
    \mathbf{I}_\beta(x|y) = \kappa_0 e^{i \frac{\omega}{\beta}|x-y|}+\frac{i \pi\omega}{\tilde{d}^2}\sum\limits_{\sigma=\pm} \kappa_\sigma \;g_{\beta v_\sigma}(x|y)\\ -\frac{i \pi\omega}{\tilde{d}^2}\sum\limits_{\sigma,g=\pm} \mathbf{A}_{\sigma g} (y) f_{\beta ,gv_\sigma}(x) \,,
}
\aleq{\label{Jbeta}
    \mathbf{J}_\beta (x|y)= \kappa_0 e^{i \frac{\omega}{\beta}(x+y)}\mathbf{B}+\frac{i \pi\omega}{\tilde{d}^2}e^{i \frac{\omega}{\beta}x}\sum\limits_{\sigma=\pm} \kappa_\sigma \;g_{\beta v_\sigma}(0|y) \mathbf{B} \\-\frac{i \pi\omega}{\tilde{d}^2}e^{i \frac{\omega}{\beta}x}\sum\limits_{\sigma,g=\pm} \mathbf{B}\mathbf{A}_{\sigma g} (y) f_{\beta,g v_\sigma}(0) \,,
}
where we used the   notation
\aleq{
    &g_{\beta v_\sigma}(x|y)=\int dz \;e^{i \frac{\omega}{\beta}|x-z|}e^{i \frac{\omega}{v_{\sigma}}|z-y|}
    =\theta_1^{\beta_1\beta_2} e^{i\frac{\omega}{\beta_1}|x-y|}\\
    &+\theta_2^{\beta_1\beta_2} e^{i\frac{\omega}{\beta_2}|x-y|}
    +\theta_3^{\beta_1\beta_2}e^{i\omega\left(\frac{x}{\beta_1}+\frac{y}{\beta_2} \right)}+\theta_4^{\beta_1\beta_2}e^{-i\omega\left(\frac{x}{\beta_1}+\frac{y}{\beta_2} \right)}
}
with coefficients
\aleq{
    \theta^{\beta_1\beta_2}_1=\frac{2 i \beta_1^2 \beta_2}{\omega (\beta_1^2-\beta_2^2)}, \quad \theta_2^{\beta_1\beta_2}=\theta_1^{\beta_2\beta_1}=-\frac{\beta_2}{\beta_1}\theta_1^{\beta_1\beta_2},\\
    \theta_3^{\beta_1\beta_2}=-\frac{i \beta_1 \beta_2}{\omega (\beta_1+\beta_2)},\quad \theta_4^{\beta_1\beta_2}=\theta_3^{\beta_1\beta_2} e^{i\omega L\left(\frac{1}{\beta_1}+\frac{1}{\beta_2} \right)}\,.
}
In these calculations we set the lower limit of integration to zero $a=0$ because all diagrams are IR finite. We also introduced
\aleq{
    &f_{\beta_1 \beta_2}(x)=\int dz \;e^{i \frac{\omega}{\beta_1}|x-z|}e^{i \frac{\omega}{\beta_2}z}=
     \theta_2^{\beta_1\beta_2} e^{i\frac{\omega}{\beta_2}x}\\
     &+\theta^{\beta_1,-\beta_2}_3 e^{i\frac{\omega}{\beta_1}x}+\theta^{\beta_1\beta_2}_4 e^{-i\frac{\omega}{\beta_1}x}\,.
}
 The last remaining integral in \eqref{integrals} reads as
\aleq{
    \mathbf{Q}(x|y)= \int dz\;\Big( g_{cv}(x|z)+\mathbf{B} e^{i\frac{\omega}{c}x}f_{vc}(z) \Big)\mathbf{C}(z|y)\\
    = \theta^{cv}_1 \mathbf{I}_c(x|y) +\theta^{cv}_2 \mathbf{I}_{v}(x|y)+ e^{i \frac{\omega}{c}x} \theta_1^{vc}\mathbf{J}_{c}(0|y)\\+ e^{i \frac{\omega}{c}x}\theta_4^{vc}  \mathbf{J}_{-v}(0|y)+e^{i \frac{\omega}{c}x}\left[ \theta_3^{cv} \mathbf{B}^{-1}+\theta_3^{v,-c}  \right] \mathbf{J}_v(0|y)\\+e^{-i \frac{\omega}{c}x} \theta_4^{cv} \mathbf{B}^{-1}\mathbf{J}_{-v}(0|y)\,,
}
where we   used   the property $g_{\beta_1\beta_2}(0|y)=f_{\beta_2\beta_1}(y)$.

Now we are ready to combine all contributions originating from integrals $\mathbf{J}$, $\mathbf{I}$ and $\mathbf{Q}$ together in the integral equation \eqref{WH_equation_compact1}.

Prefactors before the exponents with modified velocities $v_\pm$ $e^{g i\frac{\omega}{v_\sigma}x}$ are exactly zero. The only contributions that remain intact in the equation \eqref{WH_equation_compact1} are related to exponents $e^{\pm i\frac{\omega}{v}x}$ and $e^{\pm i\frac{\omega}{c}x}$.

By matching factors in front of exponents $e^{\pm i\frac{\omega}{v}x}$ we obtain two equations for matrices $\mathbf{A}_{\sigma g}$
\aleq{\label{12_equations_A}
    \sum\limits_{\sigma,g=\pm} \theta_3^{v,-gv_\sigma} \mathbf{A}_{\sigma g} (y) &= \sum\limits_{\sigma=\pm}\kappa_\sigma \theta_{3}^{vv_\sigma}e^{i\frac{\omega}{v_\sigma}y}\mathbf{1}  \,,\\
    \sum\limits_{\sigma,g=\pm} \theta_4^{v,gv_\sigma}\mathbf{A}_{\sigma g} (y)  &=
   \sum\limits_{\sigma=\pm}\kappa_\sigma \theta_{4}^{vv_\sigma}e^{-i\frac{\omega}{v_\sigma}y} \mathbf{1}   \,.
} 

Similarly, in case of $e^{\pm i\frac{\omega}{c}x}$ we obtain two additional equations 
\aleq{\label{34_equations_A}
\sum\limits_{\sigma,g=\pm} \theta_3^{c,-gv_\sigma}\mathbf{A}_{\sigma g} (y) &= \sum\limits_{\sigma=\pm}\kappa_\sigma \theta_{3}^{cv_\sigma}e^{i\frac{\omega}{v_\sigma}y}\mathbf{1}-\mathbf{T}_1(y)
\,, 
\\ \sum\limits_{\sigma,g=\pm} \theta_4^{c,gv_\sigma}\mathbf{A}_{\sigma g} (y) &=
    \sum\limits_{\sigma=\pm}\kappa_\sigma \theta_{4}^{cv_\sigma}e^{-i\frac{\omega}{v_\sigma}y}\mathbf{1} -T_2(y)\mathbf{1} \,,
}
where we introduced quantities $\mathbf{T_1(y)}$ and $T_2(y)$ which should be evaluated explicitly in terms of matrices $\mathbf{A}$. The rest of this Appendix is devoted to simplification of Eqs.~\eqref{34_equations_A}. 

The contribution for the first equation in \eqref{34_equations_A} that we need to calculate is given by two terms
\begin{equation}
    \mathbf{T}_1(y)=\frac{i \tilde{d}^2 c(v-c)}{2\pi v^2 \omega} \mathbf{B}^{-1}\mathbf{J}_{v}(0|y)+\mathbf{T}^{(1)}_1(y)  \,,
\end{equation}
where the first term is proportional to the identity matrix (check the definition \eqref{Jbeta}). In contrast, the second term is linear in $\mathbf{B}$ and defined as
\aleq{
 \mathbf{T}^{(1)}_1(y)=\frac{i\tilde{d}^2/\pi}{\omega\theta_1^{cv} -2i v } \Big[ \theta_2^{vc} \mathbf{J}_c(0|y) +\theta_3^{v,-c} \mathbf{J}_v(0|y)\\
 +\theta_4^{vc} \mathbf{J}_{-v}(0|y)+2\pi i v^2 e^{i\frac{\omega}{c}y}\mathbf{B}/\omega-2iv \mathbf{J}_c(0|y)/\omega \Big]  \,.
}
For the last term in the second equation in \eqref{34_equations_A} we obtain
\begin{equation}\label{T_2(y)}
    T_2(y) = \frac{i \tilde{d}^2 c(v-c)}{2v^2\pi \omega} e^{i\omega L\left( \frac{1}{c}+\frac{1}{v}\right)}\mathbf{B}^{-1}\mathbf{J}_{-v}(0|y)\,.
\end{equation}

We can further  simplify $\mathbf{J}_{\beta}$ by making use of the first two equations \eqref{12_equations_A}
 as follows
\aleq{\label{trick_1}
    \sum\limits_{\sigma,g=\pm}\mathbf{A}_{\sigma g} (y) f_{v,gv_\sigma}(0)=\sum\limits_{\sigma,g=\pm}\theta_2^{vv_\sigma}\mathbf{A}_{\sigma g} (y) \\+ \sum\limits_{\sigma=\pm}\kappa_\sigma \theta_{3}^{vv_\sigma}e^{i\frac{\omega}{v_\sigma}y} +\sum\limits_{\sigma=\pm}\kappa_\sigma \theta_{4}^{vv_\sigma}e^{-i\frac{\omega}{v_\sigma}y}\,.
}
We notice that $f_{-\beta_1,\beta_2}(0)=e^{-i\frac{\omega}{\beta_1}L}f_{\beta_1\beta_2}(L)$ and  obtain
\aleq{\label{trick_2}
    \sum\limits_{\sigma,g=\pm} \mathbf{A}_{\sigma g} (y) f_{-v,gv_\sigma}(0)=\sum\limits_{\sigma,g=\pm}\theta_2^{v,v_\sigma}\mathbf{A}_{\sigma g} (y) e^{ i\omega L\left(\frac{g}{v_\sigma}-\frac{1}{v}\right)}
    \\
   + \sum\limits_{\sigma=\pm}\kappa_\sigma \theta_{3}^{vv_\sigma}e^{i\frac{\omega}{v_\sigma}y} +e^{-2i\frac{\omega}{v}L}\sum\limits_{\sigma=\pm}\kappa_\sigma \theta_{4}^{vv_\sigma}e^{-i\frac{\omega}{v_\sigma}y}\,.
}
Substituting \eqref{trick_1} and \eqref{trick_2} into \eqref{Jbeta} one can show that
\begin{equation}\label{J_v_trick}
    \frac{i\tilde{d}^2}{\pi\omega}\mathbf{B}^{-1}\mathbf{J}_v(0|y)=\sum\limits_{\sigma,g=\pm}\theta_2^{vv_\sigma} \mathbf{A}_{\sigma g}(y)-\sum\limits_\sigma \kappa_\sigma \theta_2^{vv_\sigma}e^{i\frac{\omega}{v_\sigma}y}
\end{equation}
and consequently
\aleq{
     \frac{i\tilde{d}^2}{\pi\omega}\mathbf{B}^{-1}\mathbf{J}_{-v}(0|y)=e^{-i \frac{\omega}{v}L}\sum\limits_{\sigma,g=\pm}\theta_2^{vv_\sigma} \mathbf{A}_{\sigma g}(y)e^{gi \frac{\omega}{v_\sigma}L}\\
     -e^{-i \frac{\omega}{v}L}\sum\limits_\sigma \kappa_\sigma \theta_2^{v v_\sigma}e^{i\frac{\omega}{v_\sigma}(L-y)}\,.
}

These two relations allow us to simplify $T_2(y)$ defined in \eqref{T_2(y)}. 
Rearranging all terms in the second equation in \eqref{34_equations_A} we represent it the   form
\begin{equation}\label{3_equation_A}
     \sum\limits_{\sigma,g=\pm} \mathbf{A}_{\sigma g} (y) \varphi_g^\sigma=
    \sum\limits_{\sigma=\pm}\kappa_\sigma \varphi_+^\sigma e^{-i\frac{\omega}{v_\sigma}y}\,,
\end{equation}
where we introduced $ \varphi_g^\sigma =\theta_4^{c, g v_\sigma}+\frac{c(v-c)}{2 v^2} \theta_2^{vv_\sigma}e^{i \omega L \left(\frac{1}{c}+ \frac{g}{v_\sigma} \right)}$.

Finally, we are ready to address the first equation in \eqref{34_equations_A}. All terms containing $e^{i\frac{\omega}{c}y}$ are exactly canceled there. The rest of $\mathbf{T}_1^{(1)}$ can be simplified with the use of  \eqref{trick_1} and \eqref{trick_2} so that we arrive to the following compact representation
\aleq{\label{T_1(y)}
     \frac{i \omega v}{c}\mathbf{T}^{(1)}_1(y) = \sum\limits_{\sigma g}  \left[ \frac{c^2}{v^3}\theta_1^{vv_\sigma} -\varphi_g^\sigma e^{-i\omega L \left(\frac{1}{c}+\frac{g}{v_\sigma}\right)}\right] \mathbf{B}\mathbf{A}_{\sigma g}(y)\\- \sum\limits_{\sigma g}\kappa_\sigma \left[ \frac{c^2}{v^3}\theta_1^{vv_\sigma} -\varphi_-^\sigma e^{-i\omega L \left(\frac{1}{c}-\frac{1}{v_\sigma}\right)}\right] \mathbf{B} e^{i \frac{\omega}{v_\sigma}y}\,.
}
Now we are in a position to combine the results of this section together. 
For clarity we redefine all coefficients as $\tilde{\chi}_1^{\beta_1\beta_2}=-i\omega \theta_3^{\beta_1,-\beta_2}$, $\tilde{\chi}_2^{\beta_1\beta_2}=-i\omega \theta_4^{\beta_1\beta_2}$ and $\phi^\sigma_{g+}=-i\omega \varphi_g^\sigma$. 
After a series of straightforward algebraic manipulations equations \eqref{12_equations_A}, \eqref{3_equation_A}, \eqref{34_equations_A} and \eqref{T_1(y)} can be reduced to Eqs.~\eqref{Full_Eq1} in the main text.

\section{Full form of the solution for $b^{sg}_{\sigma j}$}\label{b_solution}
In this section we present the explicit form of all matrix elements $b^{sg}_{\sigma j}$. They are classified by four distinctive sectors which do not mix with each other in the integral equations \eqref{WH_1} and \eqref{WH_2}. 
We first  describe our notation and then define all sectors.  

The common factor appearing in all matrix elements has the   form
\aleq{
    X = e^{i \omega L\left(-\frac{1}{v_{-}}+\frac{1}{v_{+}}\right)} \overline{F}_{1} \overline{\tilde{F}}_{2}-e^{i \omega L\left(\frac{1}{v_{-}}+\frac{1}{v_{+}}\right)} F_{1} \tilde{F}_{2}\\
    -e^{i \omega L\left(-\frac{1}{v_{-}}-\frac{1}{v_{+}}\right)} \overline{\tilde{F}}_{1} \overline{F}_{2}+e^{i \omega L\left(\frac{1}{v_{-}}-\frac{1}{v_{+}}\right)} \tilde{F}_{1} F_{2}+F_{0}  \,. 
}
Four quantities $F_i$ are defined as 
\begin{eqnarray*} 
F_{1} &=& 2 v^2 \sum\limits_\sigma \sigma \tilde{\chi}_{1}^{v,-v_{-\sigma}}\Big(\tilde{\chi}_{1}^{c,-v_{\sigma}}-B \tilde{\chi}_{1}^{c,v_{\sigma}} \Big)\\& &+c\sum\limits_\sigma \sigma \tilde{\theta}_{2}^{v,v_{-\sigma}}\tilde{\chi}_{1}^{v,-v_{\sigma}}\Big((B+1)c+(B-1)v\Big)  \,,
\\
F_{2} &=& 2v^2\sum\limits_\sigma \sigma \tilde{\chi}_{1}^{v,\sigma v_{\sigma}} \tilde{\chi}_{1}^{c,-\sigma v_{-\sigma}}+c\sum\limits_\sigma \sigma \tilde{\theta}_{2}^{v v_{\sigma}} \tilde{\chi}_{1}^{v,-\sigma v_{-\sigma}}(c-v) \,,
\\
F_{3} &=& 2v^2\sum\limits_\sigma \sigma \tilde{\chi}_1^{c,\sigma v_+}\Big( \tilde{\chi}_1^{v,\sigma v_+}B+ \tilde{\chi}_1^{v,-\sigma v_+}\Big) \\& &
+c\tilde{\theta}_2^{vv_+}\sum\limits_g g \tilde{\chi}_1^{v,g v_+}\Big((B+1)c+(B-1)v\Big) \,, 
\\
F_0 &=&\frac{8 c^3 v^3 v_+ v_-  \tilde{d}^2}{\alpha \tilde{\chi}_1^{v,c}} \left(\frac{v^2-c^2}{gv^2 -c^2 \tilde{g}}\right)^2 ((B+1)c +(B-1)v) \,,
\end{eqnarray*}
where $\tilde{\theta}_2^{\beta_1\beta_2}=-i\omega \theta_2^{\beta_1\beta_2}$, and $B$ corresponds to one of the eigenvalues of the $\mathbf{B}$ matrix, i.e. it can be either $\rho+\tau$ or $\rho-\tau$. 

In addition, we defined two operations acting on quantities $F_i$ 
\begin{equation}
    \overline{F}_{i}=\left.F_{i}\right|_{\tilde{\chi}_{1}^{\beta, v_{-}} \leftrightarrow \tilde{\chi}_{1}^{\beta,-v_{-}}}, \quad \tilde{F}_{i}=\left.F_{i}\right|_{\tilde{\chi}_{1}^{\beta, v_{+}} \leftrightarrow \tilde{\chi}_{1}^{\beta,-v_{+}}} \,. 
\end{equation}

The first sector corresponds to $g=+, s=+$, and the matrix elements have the following form
\begin{eqnarray*}
Xb_{++}^{++} &=&e^{i \omega L\left(\frac{1}{v_{-}}-\frac{1}{v_{+}}\right)} F_{1} F_{2}-e^{i \omega L\left(-\frac{1}{v_{-}}-\frac{1}{v_{+}}\right)} \overline{F}_{1} \overline{F}_{2}\,,\\
Xb_{+-}^{++}&=&e^{i \omega L\left(-\frac{1}{v_{-}}+\frac{1}{v_{+}}\right)} \overline{F}_{1} \overline{\tilde{F}}_{2}-e^{i \omega L\left(\frac{1}{v_{-}}+\frac{1}{v_{+}}\right)} F_{1} \tilde{F}_{2}-\\
& &-F_{3}\left.F_{2}\right|_{v_{+} \rightarrow v_{-}}\,,\\
Xb_{-+}^{++}&=&-e^{i \omega L\left(-\frac{1}{v_{-}}-\frac{1}{v_{+}}\right)} F_{3}\overline{F}_{2} -F_{1}\left.F_{2}\right|_{v_{-} \rightarrow v_{+}}\,,\\
Xb_{--}^{++}&=&e^{i \omega L\left(\frac{1}{v_{-}}-\frac{1}{v_{+}}\right)} \overline{F_{3}}F_{2}+\overline{F}_{1}\left.F_{2}\right|_{v_{-} \rightarrow v_{+}}\,.
\end{eqnarray*}
The second sector is represented by the choice of indices $g=-, s=+$, and the matrix elements are defined as
\begin{eqnarray*}
Xb_{++}^{+-} &=&e^{i \omega L\left(-\frac{1}{v_{-}}+\frac{1}{v_{+}}\right)} \overline{F}_{1} \overline{\tilde{F}}_{2}-e^{i \omega L\left(\frac{1}{v_{-}}+\frac{1}{v_{+}}\right)} F_{1} \tilde{F}_{2}-\\
& &-\left.F_{3}\right|_{v_{+} \rightarrow v_{-}}\left.F_{2}\right|_{v_{-} \rightarrow v_{+}}\,,\\
Xb_{+-}^{+-}&=&-e^{i \omega L\left(-\frac{1}{v_{-}}+\frac{1}{v_{+}}\right)} \overline{\tilde{F}}_{1} \overline{\tilde{F}}_{2}+e^{i \omega L\left(\frac{1}{v_{-}}+\frac{1}{v_{+}}\right)} \tilde{F}_{1} \tilde{F}_{2}\,,\\
Xb_{-+}^{+-}&=&e^{i \omega L\left(-\frac{1}{v_{-}}+\frac{1}{v_{+}}\right)}\left.F_{3}\right|_{v_{-} \rightarrow v_{+}} \overline{\tilde{F}}_{2}+\tilde{F}_{1}\left.F_{2}\right|_{v_{-} \rightarrow v_{+}}\,,\\
Xb_{--}^{+-}&=&-e^{i \omega L\left(\frac{1}{v_{-}}+\frac{1}{v_{+}}\right)}\left.F_{3}\right|_{v_{-} \rightarrow v_{+}} \tilde{F}_{2}-\overline{\tilde{F}}_{1}\left.F_{2}\right|_{v_{-} \rightarrow v_{+}}.
\end{eqnarray*}
The third sector is defined by indices $g=+, s=-$. The corresponding matrix elements are given by
\begin{eqnarray*}
X b_{++}^{-+} &=&e^{i \omega L\left(-\frac{1}{v_{-}}-\frac{1}{v_{+}}\right)}\left.F_{3}\right|_{v_{+} \rightarrow v_{-}} \overline{F}_{2}+F_{1}\left.F_{2}\right|_{v_{+} \rightarrow v_{-}}\,,\\
X b_{+-}^{-+} &=&-e^{i \omega L\left(-\frac{1}{v_{-}}+\frac{1}{v_{+}}\right)}\left.F_{3}\right|_{v_{+} \rightarrow v_{-}} \overline{\tilde{F}}_{2}-\tilde{F}_{1}\left.F_{2}\right|_{v_{+} \rightarrow v_{-}}\,,\\
X b_{-+}^{-+}&=&e^{i \omega L\left(-\frac{1}{v_{-}}+\frac{1}{v_{+}}\right)} F_{1} \overline{\tilde{F}}_{2}-e^{i \omega L\left(-\frac{1}{v_{-}}-\frac{1}{v_{+}}\right)} \tilde{F}_{1} \overline{F}_{2}\,,\\
X b_{--}^{-+}&=&-e^{i \omega L\left(\frac{1}{v_{-}}+\frac{1}{v_{+}}\right)} F_{1} \tilde{F}_{2}+e^{i \omega L\left(\frac{1}{v_{-}}-\frac{1}{v_{+}}\right)} \tilde{F}_{1} F_{2}-\\
& &-\left.F_{3}\right|_{v_{+} \rightarrow v_{-}}\left.F_{2}\right|_{v_{-} \rightarrow v_{+}}\,.
\end{eqnarray*}
Finally, the last sector for matrix elements with $g=+, s=-$ has the form
\begin{eqnarray*}
X b_{++}^{--}&=&-e^{i \omega L\left(\frac{1}{v_{-}}-\frac{1}{v_{+}}\right)}\left.F_{3}\right|_{v_{+} \rightarrow v_{-}} F_{2}-\overline{F}_{1}\left.F_{2}\right|_{v_{+} \rightarrow v_{-}}\,,\\
X b_{+-}^{--}&=&e^{i \omega L\left(\frac{1}{v_{-}}+\frac{1}{v_{+}}\right)}\left.F_{3}\right|_{v_{+} \rightarrow v_{-}} \tilde{F}_{2}+\overline{\tilde{F}}_{1}\left.F_{2}\right|_{v_{+} \rightarrow v_{-}}\,,\\
X b_{-+}^{--}&=&-e^{i \omega L\left(\frac{1}{v_{-}}+\frac{1}{v_{+}}\right)} F_{1} \tilde{F}_{2}+e^{i \omega L\left(\frac{1}{v_{-}}-\frac{1}{v_{+}}\right)} \tilde{F}_{1} F_{2}\\
& &-\left.F_{3}\right|_{v_{-} \rightarrow v_{+}}\left.F_{2}\right|_{v_{+} \rightarrow v_{-}}\,,\\
X b_{--}^{--}&=&e^{i \omega L\left(\frac{1}{v_{-}}+\frac{1}{v_{+}}\right)} \overline{F}_{1} \tilde{F}_{2}-e^{i \omega L\left(\frac{1}{v_{-}}-\frac{1}{v_{+}}\right)} \overline{\tilde{F}}_{1} F_{2}\,.
\end{eqnarray*}

\end{document}